\documentclass[aps,floatfix,amsmath,nofootinbib,amssymb,superscriptaddress]{revtex4-2}

\usepackage{overpic}
\usepackage{amssymb}
\usepackage{indentfirst}
\usepackage{feynmf}   
\usepackage{slashed}  
\usepackage{cases}
\usepackage{color}
\usepackage{multirow}
\usepackage{epstopdf}
\usepackage{graphicx,color,bm}
\usepackage{epstopdf}

\usepackage[colorlinks,
            citecolor=green,
            anchorcolor=red,
            menucolor=red,
            linkcolor=red,
            filecolor=red,
            runcolor=red,
            urlcolor=blue,
            frenchlinks=red]{hyperref}

\begin{document}

\title{Chiral-even twist-3 GPDs for the proton in a spectator diquark model}

\author{Chentao Tan}\affiliation{School of Physics, Southeast University, Nanjing
211189, China}

\author{Zhun Lu}
\email{zhunlu@seu.edu.cn}
\affiliation{School of Physics, Southeast University, Nanjing 211189, China}

\begin{abstract}

We investigate the chiral-even twist-3 generalized parton distributions (GPDs) of valence quarks in the proton at nonzero skewness $\xi$, using a spectator model with scalar and axial-vector diquarks. We consider the exponential form factor for the nucleon-quark-diquark vertex and the axial-vector diquark with light-cone transverse polarization. We analyze the dependence of GPDs on the longitudinal momentum fraction $x$ at different $\xi$, and on the square of the transverse momentum transfer $\bm{\Delta}^2_T$ at different $x$. Our numerical results reveal distinct discontinuities in all twist-3 GPDs except $G_1$ and $\tilde{G}_1$. 
By taking the forward limit, we obtain the twist-3 parton distribution function $g_T$, which encodes the transverse spin distribution of quarks. We also compare the kinetic orbital angular momentum and the spin-orbit correlations of quarks defined by the twist-2 and twist-3 GPDs, respectively.

\end{abstract}

\maketitle

\section{Introduction}

A depiction of the internal structure of hadrons complementary to transverse momentum dependent distributions (TMDs) can be obtained by decomposing its global properties into corresponding quark and gluon quantities in transverse position space via sum rules. 
The relevant distributions are the Fourier transforms of generalized parton distributions (GPDs)~\cite{Muller:1994ses,Ji:1996ek,Radyushkin:1996nd,Radyushkin:1996ru}. 
Theoretically, they appear in the parameterizations for off-forward matrix elements of bilocal operators. 
Experimentally, they can be extracted from hard exclusive deep inelastic scattering experiments such as deeply virtual Compton scattering (DVCS) \cite{Ji:1996nm,Collins:1998be} and deeply virtual meson production (DVMP) \cite{Ji:1998pc,Goeke:2001tz,Collins:1996fb}. 
In recent years, numerous measurements have been conducted at HERA (H1 \cite{H1:2001nez,H1:2005gdw,H1:2007vrx,H1:2009wnw}, ZEUS \cite{ZEUS:2003pwh,ZEUS:2008hcd}) and HERMES \cite{HERMES:2012gbh,HERMES:2012idp}), at JLab (CLAS \cite{CLAS:2007clm,CLAS:2008ahu,Niccolai:2012sq,CLAS:2015bqi,CLAS:2015uuo,CLAS:2018bgk,CLAS:2018ddh,CLAS:2021gwi} and Hall A \cite{JeffersonLabHallA:2006prd,JeffersonLabHallA:2007jdm,JeffersonLabHallA:2012zwt,Georges:2017xjy,JeffersonLabHallA:2022pnx}), and at COMPASS \cite{Kumericki:2016ehc}. 
More accurate data could be obtained in future experiments at the Electron-Ion Collider (EIC) \cite{AbdulKhalek:2021gbh} and EicC \cite{Anderle:2021wcy}. 
Therefore, as a precise probe for the three-dimensional structure inside hadrons, GPDs have provided much information about the momentum, angular momentum, spatial distributions, and other properties of partons~\cite{Ji:1996ek,Burkardt:2000za,Bondarenko:2002pp,Burkardt:2002hr,
Ralston:2001xs,Diehl:2002he}.

Similar to PDFs and TMDs, GPDs can also be classified according to their twist \cite{Jaffe:1996zw,Meissner:2009ww}. Twist indicates the order of $1/Q$ at which a matrix element contributes to the scattering amplitude, with $Q$ being the four-momentum transfer of a given physical process. Research on GPDs has attracted considerable attention, particularly concerning the twist-2 GPDs ~\cite{Boffi:2007yc,Kumericki:2009uq,Burkardt:2002ks,Bhattacharya:2018zxi,
Bhattacharya:2019cme}, since the contributions to matrix elements are dominated by twist-2 operators in the Bjorken limit \cite{Bjorken:1968dy}. 
However, considering the energy scales in current experiments, the effects of higher-twist GPDs cannot be neglected, motivating our focus on twist-3 GPDs in this work, whose effects are generally presumed to be significant~\cite{Guo:2022cgq,Anikin:2009hk,Freund:2003qs,Kivel:2003jt,
Kiptily:2002nx,Radyushkin:2001fc,Belitsky:2001yp,Kivel:2000fg,Belitsky:2000vk,
Radyushkin:2000ap,Kivel:2000cn,Radyushkin:2000jy,Kivel:2000rb,Belitsky:2000vx}. 

Extracting GPDs from experimental data is widely recognized as a challenging task~\cite{Bertone:2021yyz,Moffat:2023svr}, particularly for twist-3 GPDs. 
Nevertheless, in addition to providing important corrections to twist-2 amplitudes in hard exclusive reactions and enabling more reliable extraction of twist-2 GPDs, studying twist-3 GPDs offers several key motivations. 
Firstly, the kinetic orbital angular momentum (OAM) of quarks can be expressed as the $x$-moment of the twist-3 GPD $G_2$~\cite{Ji:1996ek,Belitsky:2000vx,Penttinen:2000dg}. Secondly, a nontrivial relation exists between quark spin-orbit correlations and the linear combination of the $x$-moments of the twist-3 GPDs $\tilde{G}_2$ and $\tilde{G}_4$~\cite{Lorce:2014mxa,Bhoonah:2017olu}. 
Thirdly, the forward limit of the twist-3 GPD $H_{2T}^\prime$ provides information about the average transverse color Lorentz force acting on quarks in a transversely polarized nucleon~\cite{Burkardt:2008ps,Aslan:2019jis}. 
Finally, based on certain relations~\cite{Rajan:2016tlg,Rajan:2017cpx} between twist-3 GPDs and Generalized Transverse Momentum Dependent Parton Distributions (GTMDs)~\cite{Meissner:2009ww,Lorce:2013pza}, constraints can be imposed on the latter through the former. 
These motivations have prompted lattice QCD calculations for twist-3 GPDs~\cite{Dodson:2021rdq,Bhattacharya:2023nmv}. Despite their inherent challenges, lattice calculations have offered valuable insights into the understanding of nucleon structure.

In this paper, we calculate the chiral-even twist-3 GPDs of $u$ and $d$ quarks in the proton at $\xi\neq0$ using a spectator diquark model, in which the form factor of the nucleon-quark-diquark vertex is chosen as exponential, and the polarization sum of the axial-vector diquark only contains the physical polarization states. 
The model results indicate that all the twist-3 GPDs except $G_1$ and $\tilde{G}_1$ (or $\tilde{H}_{2T}$ and $\tilde{H}_{2T}^\prime$) exhibit discontinuities at $x=\pm\xi$, consistent with the conclusion in Ref.~\cite{Aslan:2018tff}. 
These two particular points $x=\pm\xi$ correspond to the region where the longitudinal momentum components of incoming or outgoing quarks in the matrix element vanish. 
However, these discontinuities neither lead to divergence of scattering amplitudes nor endanger the factorization of the DVCS amplitudes at the twist-3 accuracy~\cite{Collins:1989gx}, as the linear combinations of twist-3 GPDs entering the scattering amplitudes have been shown to precisely cancel discontinuities \cite{Aslan:2018zzk}. Consequently, it is impractical to phenomenologically address individual twist-3 GPDs via corresponding DVCS observables; 
instead, only the linear combinations of vector and axial-vector GPDs can be accessed. Therefore, to extract a single twist-3 GPD, it is necessary to investigate other hard exclusive processes where the discontinuities of GPDs do not occur~\cite{Kivel:2000rb,Aslan:2018zzk}, such as double DVCS~\cite{Zhao:2019bzg,Deja:2023tuc,Deja:2023aug}.

The paper is organized as follows: In Sec.~\ref{Sec:2}, we present two parameterizations of the chiral-even twist-3 quark GPDs and their relations. 
In Sec.~\ref{Sec:3}, we analytically calculate these GPDs in diquark model containing scalar and axial-vector diquarks. 
In Sec.~\ref{Sec:4}, numerical results for GPDs of $u$ and $d$ quarks contributed by the vector and axial-vector diquarks are shown, the corresponding physical observables are discussed, and the result for the twist-3 PDF $g_T$ defined by the forward limits of specific twist-3 GPD is presented. 
In Sec. \ref{Sec:5}, some conclusions are drawn.

\section{Definitions of chiral-even twist-3 GPDs}\label{Sec:2}

In this section, we present two definitions of the chiral-even twist-3 GPDs and discuss the relations between them. 
These GPDs provide a three-dimensional mapping of the internal structure of hadrons, characterized by the longitudinal momentum fraction $x$ of the active quark, the skewness $\xi$, and the momentum transfer squared $t = \Delta^2$. 
For spin-$\frac{1}{2}$ hadrons like the proton, the light-cone correlation function parameterized by quark GPDs is defined as the off-forward matrix element of a bilocal operator \cite{Meissner:2009ww,Diehl:2003ny}:
\begin{align}
	F^{[\Gamma]}(x,\Delta;\lambda,\lambda^\prime)=\frac{1}{2}\int \frac{dz^-}{2\pi}e^{ik \cdot z} \langle p^\prime; \lambda^\prime | \bar{\psi}\left(-\frac{1}{2}z\right)\Gamma \mathcal{W}\left(-\frac{1}{2}z,\frac{1}{2}z\right)\psi\left(\frac{1}{2}z\right)|p; \lambda\rangle \bigg|_{z^+=0,\bm{z}_T=\bm{0}_T},
	\label{eq:F}
\end{align}
where $p$ ($p^\prime$) and $\lambda$ ($\lambda^\prime$) represent the momentum and helicity of the initial (final) proton, respectively. $\Gamma$ is an element of the complete basis $\{1, \gamma_5, \gamma^\mu, \gamma^\mu \gamma_5, i\sigma^{\mu\nu} \gamma_5 \}$ with $\sigma^{\mu\nu} = \frac{i}{2} [\gamma^\mu, \gamma^\nu]$, and
\begin{align}
 \mathcal{W}\left(-\frac{1}{2}z,\frac{1}{2}z\right) \bigg|_{z^+=0,\bm{z}_T=\bm{0}_T} &= \mathcal{P} \exp \left(-ig\int^{z^-/2}_{-z^-/2} dy^- A^+(0^+, y^-, \bm{0}_T) \right)
\end{align}
is the Wilson line ensuring the color gauge invariance of the correlator (Eq.~(\ref{eq:F})), where $g$ denotes the strong coupling constant. 
In the light-cone gauge ($A^+ = 0$), the Wilson line contributes unity. 
In Addition, the reference frame chosen is the symmetric frame where $\bm{P}_T = \bm{0}_T$, allowing specification of the momenta of the initial and final protons as
\begin{align} 
p&=\left((1+\xi)P^+,\frac{M^2+\frac{\bm{\Delta}_T^2}{4}}{(1+\xi)P^+},
{-\frac{\bm{\Delta}_T}{2}}\right),\\ p&^\prime=\left((1-\xi)P^+,\frac{M^2+\frac{\bm{\Delta}_T^2}{4}}
{(1-\xi)P^+},{\frac{\bm{\Delta}_T}{2}}\right),\\
\Delta&=p^\prime-p=\left(-2\xi P^+,\frac{t+\bm{\Delta}_T^2}{-4\xi P^+},\bm{\Delta}_T\right),
\end{align}
where $P = (p + p^\prime)/2$ represents the average momentum, and $M$ is the proton mass with $p^2 = p^{\prime 2} = M^2$. Since so far  it is formidable to access to the GPDs involving negative $\xi$ via known processes, we typically consider the region $0 \leq \xi \leq 1$. 
The supporting region for $x$ in GPDs is $-1 \leq x \leq 1$, where $-\xi \leq x \leq \xi$ is referred to as the ERBL region~\cite{Efremov:1978rn, Lepage:1980fj}, and $\xi \leq x \leq 1$ ($-1 \leq x \leq -\xi$) is referred to as the DGLAP region for quarks (antiquarks)~\cite{Gribov:1972ri, Lipatov:1974qm, Borah:2012ey}. 
This work focuses on the configurations where $\xi \neq 0$, therefore the twist-3 GPDs should be separately calculated in these three regions. 

For the following discussion, it is necessary to introduce
the polarization state in a generic direction $\bm{S}=(S_T^1,S_T^2,\lambda)=(\text{sin}\theta \text{cos}\phi,\text{sin} \theta \text{sin}\phi,\text{cos} \theta)$: 
\begin{align} 	|p;S\rangle=\text{cos}(\theta/2)|p;+\rangle+\text{sin}(\theta/2)e^{i\phi}|p;-\rangle
\end{align}
which denotes that the initial proton, with both longitudinal and transverse polarization, can be expressed as a superposition of states with definite light-cone helicities $|p;+\rangle|p;+\rangle$ and $|p;-\rangle$~\cite{Diehl:2005jf}. 
Similarly, $\langle p^\prime;S|$ for the final proton has a similar superposition representation. 
Thus one can establish the relationship between the general correlator $F(x,\Delta;S)$ and the correlators $F(x,\Delta;\lambda,\lambda^\prime)$ for all possible helicity combinations~\cite{Meissner:2007rx}:
\begin{align}	
F(x,\Delta;S)=&\frac{1}{2}[F(x,\Delta;+,+)+F(x,\Delta;-,-)]
+\frac{1}{2}\lambda[F(x,\Delta;+,+)-F(x,\Delta;-,-)]\nonumber\\ &+\frac{1}{2}S_T^1[F(x,\Delta;-,+)+F(x,\Delta;+,-)]
+\frac{i}{2}S_T^2[F(x,\Delta;-,+)-F(x,\Delta;+,-)],
\end{align}
which is crucial in the calculation of GPDs.

A complete parametrization of quark GPDs at different twists for a spin-$\frac{1}{2}$ hadron has been presented in Ref.~\cite{Meissner:2009ww}. 
At twist-3 level, there are eight chiral-even GPDs, which are defined by
\begin{align} F^{[\gamma^j]}(x,\Delta;\lambda,\lambda^\prime)=&\frac{M}{2(P^+)^2}
\bar{u}(p^\prime,\lambda^\prime)\bigg[i\sigma^{+j}H_{2T}(x,\xi,t)
+\frac{\gamma^+\Delta^j_T-\Delta^+\gamma^j}{2M}E_{2T}(x,\xi,t)\nonumber\\ &+\frac{P^+\Delta_T^j-\Delta^+P_T^j}{M^2}\tilde{H}_{2T}(x,\xi,t)
+\frac{\gamma^+P_T^j-P^+\gamma^j}{M}\tilde{E}_{2T}(x,\xi,t)\bigg]u(p,\lambda),
\label{eq:Fgammaj}\\
	F^{[\gamma^j\gamma_5]}(x,\Delta;\lambda,\lambda^\prime)=&-\frac{i\epsilon^{ij}_T M}{2(P^+)^2}\bar{u}(p^\prime,\lambda^\prime)\bigg[i\sigma^{+i}H_{2T}^\prime(x,\xi,t)
+\frac{\gamma^+\Delta^i_T-\Delta^+\gamma^i}{2M}E_{2T}^\prime(x,\xi,t)\nonumber\\
	&+\frac{P^+\Delta_T^i-\Delta^+P_T^i}{M^2}\tilde{H}_{2T}^\prime(x,\xi,t)
+\frac{\gamma^+P_T^i-P^+\gamma^i}{M}\tilde{E}_{2T}^\prime(x,\xi,t)\bigg]
u(p,\lambda)\label{eq:Fgammaj5},
\end{align}
where $\epsilon_T^{ij}=\epsilon^{-+ij}$ is an antisymmetric Levi-Civita tensor with $\epsilon^{-+12}=1$. The light-cone helicity spinors of the initial and final protons $u(p,\lambda)$ and $\bar{u}(p^\prime,\lambda^\prime)$ follow conventional forms \cite{Lepage:1980fj}:
\begin{align}
	u(p,+)=\frac{1}{\sqrt{2^{3/2}p^+}}\begin{pmatrix}
		&\sqrt{2}p^++m\\
		&p_T^1+ip_T^2\\
		&\sqrt{2}p^+-m\\
		&p_T^1+ip_T^2
	\end{pmatrix},~~~~
	u(p,-)=\frac{1}{\sqrt{2^{3/2}p^+}}\begin{pmatrix}
		&-p_T^1+ip_T^2\\
		&\sqrt{2}p^++m\\
		&p_T^1-ip_T^2\\
		&-\sqrt{2}p^++m
	\end{pmatrix}.
\end{align}

By identifying the chiral-even Generalized Parton Distributions (GPDs) as the vector and axial-vector ones, a different parametrization was introduced in Ref.~\cite{Aslan:2018zzk}. 
This one relates $F^\mu$ and $\tilde{F}^\mu$ as follows:
\begin{align}
	F^\mu&=\bar{u}(p^\prime)\left[P^\mu \frac{\gamma^+}{P^+}H+P^\mu\frac{i\sigma^{+\nu}\Delta_\nu}{2MP^+}E+\Delta_T^\mu\frac{1}{2M}G_1+\gamma_T^\mu(H+E+G_2)+\Delta_T^\mu\frac{\gamma^+}{P^+}G_3+i\epsilon_T^{\mu\nu}\Delta_\nu\frac{\gamma^+\gamma_5}{P^+}G_4\right]u(p),\\
	\tilde{F}^\mu&=\bar{u}(p^\prime)\left[P^\mu \frac{\gamma^+\gamma_5}{P^+}\tilde{H}+P^\mu\frac{\Delta^+\gamma_5}{2MP^+}\tilde{E}+\Delta_T^\mu\frac{\gamma_5}{2M}(\tilde{E}+\tilde{G}_1)+\gamma^\mu_T\gamma_5(\tilde{H}+\tilde{G}_2)+\Delta_T^\mu\frac{\gamma^+\gamma_5}{P^+}\tilde{G}_3+i\epsilon_T^{\mu\nu}\Delta_\nu\frac{\gamma^+}{P^+}\tilde{G}_4\right]u(p).
\end{align}	
Based on the Dirac equation~\cite{Aslan:2018zzk,Belitsky:2000vx,Belitsky:2005qn,Lorce:2017isp},
these two types of GPDs are known to be related through the relations \cite{Aslan:2018zzk,Zhang:2023xfe}: 
\begin{align}
	G_1=&2\tilde{H}_{2T},~~~~~~~~~~~~~~~~~~~~~~~~~~~~~~~~~~~~~~~~~~~~~~~~~~	G_2=-(H+E)-\frac{1}{\xi}(1-\xi^2)H_{2T}+\xi E_{2T}-\tilde{E}_{2T},\nonumber\\ G_3=&\frac{1}{2}(H_{2T}+E_{2T}),~~~~~~~~~~~~~~~~~~~~~~~~~~~~~~~~~~~~~~
G_4=\frac{1}{2\xi}H_{2T},\nonumber\\	\tilde{G}_1=&-\tilde{E}+2\tilde{H}_{2T}^\prime,~~~~~~~~~~~~~~~~~~~~~~~~~~~~~~~~~~~~~~~~
\tilde{G}_2=-\tilde{H}+(1-\xi^2)H_{2T}^\prime-\xi^2E_{2T}^\prime-
\frac{\bm{\Delta}_T^2}{2M^2}\tilde{H}_{2T}^\prime+\xi\tilde{E}_{2T}^\prime,\nonumber\\	\tilde{G}_3=&-\frac{\xi}{2}(H_{2T}^\prime+E_{2T}^\prime)
-\frac{\xi\bar{M}^2}{M^2}\tilde{H}_{2T}^\prime
+\frac{1}{2}\tilde{E}_{2T}^\prime,
~~~~~~~\tilde{G}_4=-\frac{1}{2}(H_{2T}^\prime+E_{2T}^\prime)
-\frac{\bar{M}^2}{M^2}\tilde{H}_{2T}^\prime
\label{eq:relation},
\end{align}
where $\bar{M}^2 = M^2 + t/4$. Additionally, various other parameterizations of twist-3 GPDs have been proposed in the literature \cite{Belitsky:2001yp,Aslan:2019jis,Guo:2021aik,Belitsky:2001ns,Ji:2012ba}.

\section{Twist-3 GPDs in the spectator diquark model}\label{Sec:3}

\begin{figure}
	\centering
	\includegraphics[width=0.4\columnwidth]{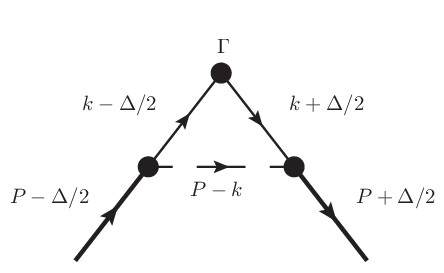}
	\caption{Kinematics for quark GPDs in the spectator diquark model.}
	\label{fig:quark}
\end{figure}

In this section, we analytically calculate the chiral-even twist-3 quark Generalized Parton Distributions (GPDs) at $\xi \neq 0$ using the  diquark model with scalar diquarks and axial-vector diquarks. 
The diquark model in different forms has been extensively utilized in studying various aspects of hadron structure, including TMDs and leading-twist GPDs of quarks.
In this model, the proton is treated as a bound state comprising an active quark with mass $m$ and a spectator diquark with mass $M_X$. 
The type of diquark (scalar or axial-vector) depends on its spin state. 
Typically, to describe the distributions of valence quarks, both types of diquarks are simultaneously involved. 
Figure (\ref{fig:quark}) illustrates the Feynman diagram at the lowest order used to compute the correlator (\ref{eq:F}), where the quark-quark vertex corresponds to the Dirac matrix $\Gamma$. The nucleon-quark-diquark vertices incorporate form factors to eliminate divergences arising from integrating the correlator over the entire transverse momentum space. For this purpose, we adopt the following forms for the vertices:
\begin{align}
	\text{nucleon-quark-scalar-diquark vertex}:~&ig_s^\pm I,\\
	\text{nucleon-quark-axial-vector-diquark vertex}:~&i\frac{g_a^\pm }{\sqrt{2}}\gamma^\mu\gamma_5,
\end{align}
where $g_X^\pm$ are chosen as the exponential form factors~\cite{Ma:2019agv,Lu:2012ez}. They take the form 
\begin{align}
	g_X^\pm\equiv g_X(\bm{k}_T\pm\frac{\bm{\Delta}_T}{2})=g_X\text{exp}
\bigg\{-\frac{1}{8\Lambda_X^2}\bigg[\frac{m^2+(\bm{k}_T\pm\frac{\bm{\Delta}_T}{2})^2}{|x|}
+\frac{M_X^2+(\bm{k}_T\pm\frac{\bm{\Delta}_T}{2})^2}{1-|x|}\bigg]\bigg\}.
\end{align}
Here, the coupling constants $g_X$ and the cutoffs $\Lambda_X$ are the free parameters of the model, and the superscript $+~(-)$ corresponds to the vertex of the outing (incoming) quark. 
The propagators of the scalar and axial-vector diquarks are given by
\begin{align}
	\text{scalar diquark propagator}:~&\frac{i}{(P-k)^2-M_s^2+i\epsilon},\\
	\text{axial-vector diquark propagator}:~&\frac{i}{(P-k)^2-M_a^2+i\epsilon}d^{\mu\nu}(P-k),
\end{align}
respectively, where the polarization tensor $d^{\mu\nu}$ sums over all polarization states of the axial-vector diquarks. 
Several choices for $d^{\mu\nu}$ have been discussed in Refs.~\cite{Bacchetta:2008af,Gamberg:2007wm,Jakob:1997wg,Bacchetta:2003rz,Brodsky:2000ii}, and we adopt the following form: 
\begin{align}
d^{\mu\nu}(P-k)=-g^{\mu\nu}+\frac{(P-k)^\mu n_-^\nu+(P-k)^\nu n_-^\mu}{(P-k)\cdot n_-}-\frac{M_a^2}{[(P-k)\cdot n_-]^2}n_-^\mu n_-^\nu
\label{eq:dmunu},
\end{align}
which only involves the light-cone transverse polarization states of the axial-vector diquark. 
Another reason for this choice is that it can give the simplest and most easily interpretable results of TMDs~\cite{Bacchetta:2008af}.

\subsection{GPDs in the spectator diquark model with scalar diquark}

We start by calculating of the twist-3 GPDs in Eqs.~(\ref{eq:Fgammaj}-\ref{eq:Fgammaj5}). According to Fig.~(\ref{fig:quark}), the correlator~(\ref{eq:F}) in the spectator diquark model with only scalar diquark can be expressed as
\begin{align}
	F^{s[\Gamma]}(x,\Delta;\lambda,\lambda^\prime)=\int \frac{dk^-d^2\bm{k}_T}{2(2\pi)^4}\frac{ig_s^+g_s^-\bar{u}(p^\prime,\lambda^\prime)
(\slashed{k}+\frac{\slashed{\Delta}}{2}+m)
\Gamma(\slashed{k}-\frac{\slashed{\Delta}}{2}+m)u(p,\lambda)}{D^s_{\text{GPD}}}
	\label{eq:Fs},
\end{align}
where
\begin{align} D^X_{\text{GPD}}=\bigg[\left(k+\frac{\Delta}{2}\right)^2-m^2+i\epsilon\bigg]
\bigg[\left(k-\frac{\Delta}{2}\right)^2-m^2+i\epsilon\bigg]
\bigg[(P-k)^2-M_X^2+i\epsilon\bigg].
\end{align}

In order to obtain the analytical results, one needs to evaluate the integration over $k^-$ as follows:
\begin{align} I \textrm{or} I_k=\int^{+\infty}_{-\infty}dk^-\frac{1 \textrm{or} k^-}{D_{\text{GPD}}}=
\frac{1}{C}\int^{+\infty}_{-\infty}dk^-\frac{1 \textrm{or} k^-}{(k^--k_1^-)(k^--k_2^-)(k^--k_3^-)}
	\label{eq:IIk},
\end{align}
where
\begin{align}
	C=-8(x+\xi)(x-\xi)(1-x)(P^+)^3.
\end{align}
The poles of the integrand in Eq.~(\ref{eq:IIk}) are given by
\begin{align} k_1^-&=-\frac{\Delta^-}{2}+\frac{(\bm{k}_T
+\frac{\bm{\Delta}_T}{2})^2+m^2-i\epsilon}{2(x-\xi)P^+},\\ k_2^-&=\frac{\Delta^-}{2}+\frac{(\bm{k}_T
-\frac{\bm{\Delta}_T}{2})^2+m^2-i\epsilon}{2(x+\xi)P^+},\\
	k_3^-&=P^--\frac{\bm{k}_T^2+M_X^2-i\epsilon}{2(1-x)P^+},
\end{align}
corresponding to the propagators of  the outgoing quark $(k+\frac{\Delta}{2})$, the incoming quark $(k-\frac{\Delta}{2})$, and the spectator diquark  $(P-k)$, respectively.
It is obvious that the position of the poles depends on $x$. 
Therefore, as mentioned in Sec.~(\ref{Sec:2}), we must integrate over $k^-$ by contour integration in the three regions of $x$. 
These integrals can be performed by using the replacements:
\begin{align}
	\frac{1}{\left(k+\frac{\Delta}{2}\right)^2-m^2+i\epsilon}&\rightarrow \frac{-2\pi i}{2(x-\xi)P^+}\delta\left(k^--k_1^-\right),\\
	\frac{1}{\left(k-\frac{\Delta}{2}\right)^2-m^2+i\epsilon}&\rightarrow \frac{-2\pi i}{2(x+\xi)P^+}\delta\left(k^--k_2^-\right),\\
	\frac{1}{(P-k)^2-M_X^2+i\epsilon}&\rightarrow \frac{-2\pi i}{2(1-x)P^+}\delta\left(k^--k_3^-\right).
\end{align}
Then $I(I_k)$ are obtained as
\begin{align}
	I(I_k) = 
	\begin{cases}
		0,&   -1\leq x\leq -\xi, \\
		-\frac{2\pi i}{C}\frac{1(k_2^-)}{(k_2^--k_1^-)(k_2^--k_3^-)}, &  -\xi \leq  x\leq \xi,\\
		\frac{2\pi i}{C}\frac{1(k_3^-)}{(k_3^--k_1^-)(k_3^--k_2^-)}, &  \xi\leq x \leq 1, 
	\end{cases}
\end{align}
where $-1\leq x\leq -\xi$ corresponds to the cut through the outgoing quark, $-\xi \leq  x\leq \xi$ corresponds to the cut through the incoming quark, and $\xi\leq x \leq 1$ corresponds to the cut through the spectator diquarks. 
We find that both $I$ and $I_k$ equal $0$ for $-1\leq x\leq -\xi$, which is due to the absence of antiquark distributions for a fermion target at order $\mathcal{O}(g^2)$, which is a general conclusion~\cite{Aslan:2018tff,Aslan:2018zzk}. 
Furthermore, the authors of Ref.~\cite{Aslan:2018zzk} also showed that the integrals $I$ and $I_k$ exhibit continuity and discontinuity at $x=\pm\xi$.  

After using the Gordon identities and evaluating $k^-$-integrals, the twist-3 GPDs contributed by the scalar diquarks can be constructed from the form
\begin{align}
	\text{GPD}(x,\xi,t) = 
	\begin{cases}
		0,&   -1\leq x\leq -\xi, \\
		\frac{g_s^2}{2(2\pi)^3}\int d^2\bm{k}_T\frac{N_i^{-\xi \leq  x\leq \xi}}{D_1D_2^{-\xi \leq  x\leq \xi}}\text{exp}\bigg\{-\frac{1}{4\Lambda_s^2}\bigg[\frac{(1-|x|)m^2
+|x|M_s^2+\bm{k}_T^2+\frac{\bm{\Delta}_T^2}{4}}{|x|(1-|x|)}\bigg]\bigg\}, &  -\xi \leq  x\leq \xi,\\
\frac{g_s^2}{(2\pi)^3}\int d^2\bm{k}_T\frac{N_i^{\xi\leq x \leq 1}}{D_1D_2^{\xi\leq x \leq 1}}\text{exp}\bigg\{-\frac{1}{4\Lambda_s^2}\bigg[\frac{(1-|x|)m^2
+|x|M_s^2+\bm{k}_T^2+\frac{\bm{\Delta}_T^2}{4}}{|x|(1-|x|)}\bigg]\bigg\}, &  \xi\leq x \leq 1, 
\end{cases}
\label{eq:twist3GPD}
\end{align}
where the denominators are given by
\begin{align}	
D_1=&(1+\xi)^2\bm{k}_T^2+\frac{1}{4}(1-x)^2\bm{\Delta}_T^2-(1-x)(1+\xi)
\bm{k}_T\cdot\bm{\Delta}_T\nonumber\\
	&+(1-x)(1+\xi)m^2+(x+\xi)(1+\xi)M_X^2-(1-x)(x+\xi)M^2,\nonumber\\
	D_2^{-\xi \leq  x\leq \xi}=&\xi(1-\xi^2)\bm{k}_T^2+\frac{1}{4}(1-x^2)\xi\bm{\Delta}_T^2
+x(1-\xi^2)\bm{k}_T\cdot\bm{\Delta}_T\nonumber\\
	&+\xi(1-\xi^2)m^2-\xi(x^2-\xi^2)M^2,\nonumber\\
	D_2^{\xi\leq x \leq 1}=&(1-\xi)^2\bm{k}_T^2+\frac{1}{4}(1-x)^2\bm{\Delta}_T^2
+(1-x)(1-\xi)\bm{k}_T\cdot\bm{\Delta}_T\nonumber\\
	&+(1-x)(1-\xi)m^2+(x-\xi)(1-\xi)M_X^2-(1-x)(x-\xi)M^2.
\end{align}
The numerators in Eq.~(\ref{eq:twist3GPD}) are listed as follows:
\begin{align}
N_{H_{2T}}^{-\xi \leq  x\leq \xi}=&N_{H_{2T}}^{\xi\leq x \leq 1}=0,\\
N_{E_{2T}}^{-\xi \leq  x\leq \xi}=&(1+\xi)\frac{\bm{k}_T\cdot\bm{\Delta}_T}{\bm{\Delta}_T^2}
\bigg[(1-\xi^2)(\bm{k}_T^2-\bm{k}_T\cdot\bm{\Delta}_T+m^2)
-\xi(x+\xi)(\bm{\Delta}_T^2+4M^2)\frac{\bm{k}_T\cdot\bm{\Delta}_T}{\bm{\Delta}_T^2}
+(1-x^2)\frac{\bm{\Delta}^2_T}{4}\nonumber\\
&-(x^2-\xi^2)M^2\bigg],\\
	N_{E_{2T}}^{\xi\leq x \leq 1}=&\frac{\bm{k}_T\cdot\bm{\Delta}_T}{\bm{\Delta}_T^2}
\bigg[-(1-\xi^2)(\bm{k}_T^2+M_s^2)-\xi(1-x)(\bm{\Delta}_T^2+4M^2)
\frac{\bm{k}_T\cdot\bm{\Delta}_T}{\bm{\Delta}_T^2} +(1-x)^2\bigg(\frac{\bm{\Delta}_T^2}{4}+M^2\bigg)\bigg],\\
	N_{\tilde{H}_{2T}}^{-\xi \leq  x\leq \xi}=&(x+\xi)(1+\xi)(1-\xi^2)M\frac{\bm{k}_T\cdot\bm{\Delta}_T}{\bm{\Delta}_T^2}\bigg(2\xi M\frac{\bm{k}_T\cdot\bm{\Delta}_T}{\bm{\Delta}_T^2}+m+xM\bigg),\\
	N_{\tilde{H}_{2T}}^{\xi\leq x \leq 1}=&(1-x)(1-\xi^2)M\frac{\bm{k}_T\cdot\bm{\Delta}_T}{\bm{\Delta}_T^2}\bigg(2\xi M\frac{\bm{k}_T\cdot\bm{\Delta}_T}{\bm{\Delta}_T^2}+m+xM\bigg),\\
	N_{\tilde{E}_{2T}}^{-\xi \leq  x\leq \xi}=&(1+\xi)\bigg\{ \frac{\bm{k}_T\cdot\bm{\Delta}_T}{\bm{\Delta}_T^2}
\bigg[x(1-\xi^2)\bm{k}_T\cdot\bm{\Delta}_T+\xi\bigg((1-\xi^2)(\bm{k}_T^2+m^2)
+(1-x^2)\frac{\bm{\Delta}_T^2}{4}-(x^2-\xi^2)M^2\bigg)\nonumber\\ &-(x+\xi)(\bm{\Delta}_T^2+4\xi^2M^2)\frac{\bm{k}_T\cdot\bm{\Delta}_T}
{\bm{\Delta}_T^2}\bigg]\nonumber\\ &-\frac{1}{2}(1-\xi)\bigg[\bigg((1+\xi)\bm{k}_T-(1-x)\frac{\bm{\Delta}_T}{2}\bigg)^2
+((1+\xi)m+(x+\xi)M)^2\bigg]\bigg\},\\
	N_{\tilde{E}_{2T}}^{\xi\leq x \leq 1}=& \frac{\bm{k}_T\cdot\bm{\Delta}_T}{\bm{\Delta}_T^2}
\bigg\{(1-x)(1-\xi^2)\bm{k}_T\cdot\bm{\Delta}_T-\xi
\bigg[(1-\xi^2)(\bm{k}_T^2+M_s^2)-(1-x)^2
\bigg(\frac{\bm{\Delta}_T^2}{4}+M^2\bigg)\bigg]\nonumber\\ &-(1-x)(\bm{\Delta}_T^2+4\xi^2M^2)\frac{\bm{k}_T\cdot\bm{\Delta}_T}
{\bm{\Delta}_T^2}\bigg\}-\frac{1}{2}(1-x)(1-\xi^2)[(m+M)^2-M_s^2],\\
N_{H_{2T}^\prime}^{-\xi\leq x \leq \xi}=&(1+\xi)\bigg\{\frac{\bm{k}_T\cdot\bm{\Delta}_T}{\bm{\Delta}_T^2}
\bigg[x(1-\xi^2)\bm{k}_T\cdot\bm{\Delta}_T+\xi
\bigg((1-\xi^2)(\bm{k}_T^2+m^2)+(1-x^2)
\frac{\bm{\Delta}_T^2}{4}-(x^2-\xi^2)M^2\bigg)\bigg]\nonumber\\
	&+\frac{1}{2M}\bigg[(1-\xi^2)((m-\xi M)\bm{k}_T^2-(m+xM)\bm{k}_T\cdot\bm{\Delta}_T+m^3+(2x+\xi)m^2M)-\xi M\bigg((1-x^2)\frac{\bm{\Delta}_T^2}{4}\nonumber\\ &-(x^2-\xi^2)M^2\bigg)+m\bigg((1-x)(1-x-2\xi)\frac{\bm{\Delta}_T^2}{4}
+(x^2+2x\xi(1-\xi)+\xi^2(1-\xi^2))M^2\bigg)\bigg]\bigg\},
\end{align}
\begin{align}	
N_{H_{2T}^\prime}^{\xi\leq x \leq1}=&\frac{\bm{k}_T\cdot\bm{\Delta}_T}{\bm{\Delta}_T^2}
\bigg\{(1-x)(1-\xi^2)\bm{k}_T\cdot\bm{\Delta}_T-\xi\bigg[(1-\xi^2)
(\bm{k}_T^2+M_s^2)-(1-x)^2\bigg(\frac{\bm{\Delta}_T^2}{4}+M^2\bigg)\bigg]\bigg\}\nonumber\\ &-\frac{1}{2M}\bigg\{(1-\xi^2)[(m+M)\bm{k}_T^2-(1-x)m^2M+(m+xM)M_s^2]+(1-x)^2(m+M)
\frac{\bm{\Delta}_T^2}{4}\nonumber\\
&-M^2[(1-x)(1+x-2\xi^2)m+(1-x)(x-\xi^2)M]\bigg\},\\
N_{E_{2T}^\prime}^{-\xi\leq x \leq \xi}=&(1+\xi)\bigg\{-\frac{\bm{k}_T\cdot\bm{\Delta}_T}{\bm{\Delta}_T^2}
\bigg[x(1-\xi^2)\bm{k}_T\cdot\bm{\Delta}_T+\xi\bigg((1-\xi^2)(\bm{k}_T^2+m^2)+(1-x^2)
\frac{\bm{\Delta}_T^2}{4}-(x^2-\xi^2)M^2\bigg)\nonumber\\ &-(x+\xi)(\bm{\Delta}_T^2+4M^2)\frac{\bm{k}_T\cdot\bm{\Delta}_T}{\bm{\Delta}_T^2}\bigg]
+\frac{1}{2}(1-\xi)\bigg[(1+\xi)(1-2x-\xi)m^2-2(x+\xi)(1+\xi)mM\nonumber\\ &+(x+\xi)(x-\xi-2)M^2+\bigg((1+\xi)\bm{k}_T-(1-x)
\frac{\bm{\Delta}_T}{2}\bigg)^2\bigg]\bigg\},\\
	N_{E_{2T}^\prime}^{\xi\leq x \leq 1}=&-\frac{\bm{k}_T\cdot\bm{\Delta}_T}{\bm{\Delta}_T^2}
\bigg\{(1-x)(1-\xi^2)\bm{k}_T\cdot\bm{\Delta}_T-\xi
\bigg[(1-\xi^2)(\bm{k}_T^2+M_s^2)-(1-x)^2
\bigg(\frac{\bm{\Delta}_T^2}{4}+M^2\bigg)\bigg]\nonumber\\ &-(1-x)(\bm{\Delta}_T^2+4M^2)\frac{\bm{k}_T\cdot\bm{\Delta}_T}
{\bm{\Delta}_T^2}\bigg\}-\frac{1}{2}(1-x)(1-\xi^2)[(m+M)^2+M_s^2],\\		
N_{\tilde{H}^\prime_{2T}}^{-\xi \leq  x\leq \xi}=&(x+\xi)(1+\xi)(1-\xi^2)M\bigg\{\frac{1-x}{2}(m+M)-
\frac{\bm{k}_T\cdot\bm{\Delta}_T}{\bm{\Delta}_T^2}\bigg[\xi(m+M) +2M\frac{\bm{k}_T\cdot\bm{\Delta}_T}{\bm{\Delta}_T^2}\bigg]\bigg\},\\
	N_{\tilde{H}^\prime_{2T}}^{\xi\leq x \leq 1}=&(1-x)(1-\xi^2)M\bigg\{\frac{1-x}{2}(m+M)-\frac{\bm{k}_T\cdot\bm{\Delta}_T}
{\bm{\Delta}_T^2}\bigg[\xi(m+M) +2M\frac{\bm{k}_T\cdot\bm{\Delta}_T}{\bm{\Delta}_T^2}\bigg]\bigg\},\\
	N_{\tilde{E}^\prime_{2T}}^{-\xi \leq  x\leq \xi}=&(1+\xi)\frac{\bm{k}_T\cdot\bm{\Delta}_T}{\bm{\Delta}_T^2}
\bigg[-(1-\xi^2)(\bm{k}_T^2-\bm{k}_T\cdot\bm{\Delta}_T+m^2)-(1-x^2)
\frac{\bm{\Delta}_T^2}{4}+\xi(x+\xi)(\bm{\Delta}_T^2+4M^2)
\frac{\bm{k}_T\cdot\bm{\Delta}_T}{\bm{\Delta}_T^2}\nonumber\\
	&+(x^2-\xi^2)M^2\bigg],\\
	N_{\tilde{E}^\prime_{2T}}^{\xi \leq  x\leq 1}=&\frac{\bm{k}_T\cdot\bm{\Delta}_T}{\bm{\Delta}_T^2}
\bigg[(1-\xi^2)(\bm{k}_T^2+M_s^2)-(1-x)^2\bigg(\frac{\bm{\Delta}_T^2}{4}
+M^2\bigg)+\xi(1-x)(\bm{\Delta}_T^2+4M^2)\frac{\bm{k}_T\cdot\bm{\Delta}_T}
{\bm{\Delta}_T^2}\bigg],
\end{align}
where we have used 
\begin{align}
	\int^{+\infty}_{-\infty}d^2\bm{k}_T\frac{k_T^\mu}{D_{\text{GPD}}}=\Delta_T^\mu \int^{+\infty}_{-\infty}d^2\bm{k}_T\frac{\bm{k}_T\cdot \bm{\Delta}_T}{\bm{\Delta}_T^2D_{\text{GPD}}},
	\label{eq:int}
\end{align}
Here the twist-3 GPD $H_{2T}$ (or $G_4$) contributed by scalar diquarks are zero.

We can also obtain the results of the vector and axial-vector twist-3 GPDs based on the relations in Eq.~(\ref{eq:relation}) and results of the twist-2 GPDs in Refs.~\cite{Bhattacharya:2018zxi,Bhattacharya:2019cme}. 
All the scalar-diquark contributions to the twist-2 GPDs present the same numerators with different coefficients in the ERBL and DGLAP regions, while the numerators of most twist-3 GPDs are significantly different in these two regions. 
It originates from the fact that the latter with $M_X$ appearing in the numerators involves the integral $I_k$, and the position of the poles depends on the value of $x$.

\subsection{GPDs in the spectator diquark model with axial-vector diquarks}

With the inclusion of axial-vector diquarks, the correlator in Eq.~(\ref{eq:F}) is written as
\begin{align}
	F^{a[\Gamma]}(x,\Delta;\lambda,\lambda^\prime)=\int \frac{dk^-d^2\bm{k}_T}{4(2\pi)^4} d_{\mu\nu}(P-k)\frac{ig_a^+g_a^-\bar{u}(p^\prime,\lambda^\prime)\gamma^\mu\gamma_5(\slashed{k}+\frac{\slashed{\Delta}}{2}+m)\Gamma(\slashed{k}-\frac{\slashed{\Delta}}{2}+m)\gamma^\nu\gamma_5u(p,\lambda)}{D^a_{\text{GPD}}}
	\label{eq:Fa}.
\end{align}
Similarly, we can obtain the model results of the twist-3 GPDs defined by the two different types of parametrization, but the explicit expressions after integrating $k^-$ will be skipped because of their complexity. 
For completeness, we also calculated the vector and axial-vector twist-2 GPDs in this model:
\begin{align}
	\text{GPD}(x,\xi,t) = 
	\begin{cases}
		0,&   -1\leq x\leq -\xi, \\
		\frac{g_a^2}{4(2\pi)^3}\int d^2\bm{k}_T\frac{N_i^{-\xi \leq  x\leq \xi}}{D_1D_2^{-\xi \leq  x\leq \xi}}\text{exp}\bigg\{-\frac{1}{4\Lambda_a^2}\bigg[\frac{(1-|x|)m^2+|x|M_a^2
+\bm{k}_T^2+\frac{\bm{\Delta}_T^2}{4}}{|x|(1-|x|)}\bigg]\bigg\}, &  -\xi \leq  x\leq \xi,\\
		\frac{g_a^2}{2(2\pi)^3}\int d^2\bm{k}_T\frac{N_i^{\xi\leq x \leq 1}}{D_1D_2^{\xi\leq x \leq 1}}\text{exp}\bigg\{-\frac{1}{4\Lambda_a^2}\bigg[\frac{(1-|x|)m^2+|x|M_a^2+\bm{k}_T^2
+\frac{\bm{\Delta}_T^2}{4}}{|x|(1-|x|)}\bigg]\bigg\}, &  \xi\leq x \leq 1, 
	\end{cases}
\end{align}
where
\begin{align}
	N_H^{-\xi\leq x\leq \xi}=&\frac{(x+\xi)(1+\xi)}{(1-x)^2}\bigg\{(1-x)(1-\xi^2)(1-x+2\xi)
\bm{k}_T^2-2(1-\xi^2)(x^2-\xi^2)M_a^2-(1-x^2)(1-x)(1+x-2\xi)
\frac{\bm{\Delta}_T^2}{4}\nonumber\\ &+(1-x)\frac{\bm{k}_T\cdot\bm{\Delta}_T}{\bm{\Delta}_T^2}
[(2x(1-\xi^2)-\xi(1-x^2))\bm{\Delta}_T^2+4\xi^3M((x^2-\xi^2)M+(1-\xi^2)m)]\nonumber\\
	&+(1-x)(1-\xi^2)(1-3x+2\xi)m^2+2(1-x)^2(1-\xi^2)(x-\xi^2)mM\nonumber\\
	&+(1-x)(x^2-\xi^2)[3-2\xi(1+\xi)-x(1-2\xi^2)]M^2\bigg\},\\
	N_H^{\xi\leq x\leq 1}=&\frac{(1-\xi^2)(1+x^2-2\xi^2)}{(1-x)}\bm{k}_T^2-(1-x)\bigg[(1+x^2-2\xi^2)
\frac{\bm{\Delta}_T^2}{4}-(1-\xi^2)m^2-2(1-\xi^2)(x-\xi^2)mM\nonumber\\
	&-(1-2\xi^2)(x^2-\xi^2)M^2\bigg]+\xi \frac{\bm{k}_T\cdot\bm{\Delta}_T}{\bm{\Delta}_T^2}\{(1+x^2-2\xi^2)\bm{\Delta}_T^2+4\xi^2 M[m+x^2M-\xi^2(m+M)]\},\\
	N_E^{-\xi\leq x\leq \xi}=&\frac{2(x+\xi)(1+\xi)(1-\xi^2)}{1-x}M\bigg\{(x^2-\xi^2)
\bigg[2\xi\frac{\bm{k}_T\cdot\bm{\Delta}_T}{\bm{\Delta}_T^2}-(1-x)\bigg]M\nonumber\\
	&+\bigg[x^2+\xi^2-x(1+\xi^2)+2\xi(1-\xi^2)\frac{\bm{k}_T\cdot\bm{\Delta}_T}
{\bm{\Delta}_T^2}\bigg]m\bigg\},\\			
	N_E^{\xi\leq x\leq 1}=&2(1-\xi^2)M\bigg\{(x^2-\xi^2)\bigg[2\xi\frac{\bm{k}_T\cdot\bm{\Delta}_T}
{\bm{\Delta}_T^2}-(1-x)\bigg]M+\bigg[x^2+\xi^2-x(1+\xi^2)+2\xi(1-\xi^2)
\frac{\bm{k}_T\cdot\bm{\Delta}_T}{\bm{\Delta}_T^2}\bigg]m\bigg\},\\
	N_{\tilde{H}}^{-\xi\leq x\leq \xi}=&\frac{(x+\xi)(1+\xi)}{(1-x)^2}\bigg\{(1-x)(1-\xi^2)(1-x+2\xi)
\bm{k}_T^2-(1-x^2)(1-x)(1+x-2\xi)\frac{\bm{\Delta}_T^2}{4}\nonumber\\	&+(1-x)\frac{\bm{k}_T\cdot\bm{\Delta}_T}{\bm{\Delta}_T^2}[2x(1-\xi^2)(\bm{\Delta}_T^2+2\xi mM)+\xi(4(x^2-\xi^2)M^2-(1-x^2)\bm{\Delta}_T^2)]-2(x^2-\xi^2)(1-\xi^2)M_a^2\nonumber\\
	&-2x(1-x)^2(1-\xi^2)mM+(1-x)(1+x-2\xi)[(x^2-\xi^2)M^2-(1-\xi^2)m^2]\bigg\},\\		
	N_{\tilde{H}}^{\xi\leq x\leq 1}=&\frac{(1-\xi^2)(1+x^2-2\xi^2)}{1-x}\bm{k}_T^2-(1-x)\bigg[(x^2-\xi^2)
\bigg(\frac{\bm{\Delta}_T^2}{4}+M^2\bigg)+(1-\xi^2)\bigg(\frac{\bm{\Delta}_T^2}{4}
+2xmM+m^2\bigg)\bigg]\nonumber\\	&+\xi\frac{\bm{k}_T\cdot\bm{\Delta}_T}{\bm{\Delta}_T^2}[(x^2-\xi^2)(\bm{\Delta}_T^2+4M^2)
+(1-\xi^2)(\bm{\Delta}_T^2+4xmM)],\\
	N_{\tilde{E}}^{-\xi\leq x\leq \xi}=&2\frac{(x+\xi)(1+\xi)(1-\xi^2)}{\xi(1-x)}M\bigg\{-\xi(1-x)^2m+2[x(1-\xi^2)m
+(x^2-\xi^2)M]\frac{\bm{k}_T\cdot\bm{\Delta}_T}{\bm{\Delta}_T^2}\bigg\},\\
	N_{\tilde{E}}^{\xi\leq x\leq 1}=&2\frac{1-\xi^2}{\xi}M\bigg\{-\xi(1-x)^2m+2[x(1-\xi^2)m+(x^2-\xi^2)M]
\frac{\bm{k}_T\cdot\bm{\Delta}_T}{\bm{\Delta}_T^2}\bigg\}.
\end{align}

We note that the twist-2 GPDs $H$ and $\tilde{H}$ with axial-vector diquark contain the integral $I_k$ due to the presence of $M_X$ in the numerators. However, this mass term is accompanied by the factor $x^2 - \xi^2$, ensuring that discontinuities do not occur. Therefore, all twist-2 GPDs exhibit continuity at $x = \pm\xi$~\cite{Bhattacharya:2018zxi,Bhattacharya:2019cme}.
In contrast, all twist-3 GPDs except $G_1$ and $\tilde{G}_1$ (or $\tilde{H}_{2T}$ and $\tilde{H}_{2T}^\prime$) are discontinuous at these points. However, the continuity of $G_1$ and $\tilde{G}_1$ may be due to the simplicity of models~\cite{Aslan:2018zzk}.

By replacing the integration variable $\bm{k}_T$ with $-\bm{k}_T$, one can verify the symmetry behavior of twist-3 GPDs under the transformation $\xi \rightarrow -\xi$. 
Notably, this transformation switches the positions of the poles of quark propagators, resulting in even denominators in both ERBL and DGLAP regions with respect to $\xi$.

Finally, we find that the GPDs $H_{2T}$, $E_{2T}$, $\tilde{H}_{2T}$, and $\tilde{E}_{2T}^\prime$ are odd functions of $\xi$, while $\tilde{E}_{2T}$, $H_{2T}^\prime$, $E_{2T}^\prime$, and $\tilde{H}_{2T}^\prime$ are even functions of $\xi$. These findings are consistent with the model-independent analysis from the hermiticity constraint in Ref.~\cite{Meissner:2009ww}.

\section{Numerical results and discussion}\label{Sec:4}

In this section, we present the numerical results of the chiral-even vector and axial-vector twist-3 GPDs of $u$ and $d$ quarks, and discuss the forward limits of certain twist-3 GPDs. 
To combine the scalar diquark and axial-vector diquark contributions, we adopt the following spin-flavor relations for the unpolarized quark distributions 
\begin{align}
	{f}_1^{u} = \frac{3}{2} {f}_1^{u(s)} + \frac{1}{2} {f}_1^{ u(a)}, ~~~~~ {f}_1^{ d} = {f}_1^{ d(a^\prime)} , \label{eq:ud}
\end{align}
where ${f}_1^{u(s)}$ is the  scalar diquark contribution to the unpolarized PDF of the $u$ valence quark, and $ {f}_1^{ u(a)}$ and  ${f}_1^{ d(a)}$ are the axial vector diquark contribution to those of the $u$ and $d$ quarks, respectively.  
The values of the parameters are adopted as~\cite{Lu:2012ez}:
\begin{align}
	m&=0.35~\textrm{GeV},~~~~~~\Lambda_{s/a}=0.33\ \textrm{GeV},\nonumber\\
    M_s&= 0.6\ \textrm{GeV},~~~~~~~~~M_{a}= 0.8\ \textrm{GeV}	,	
\end{align}
and the values of coupling constants, i.e. $g_s=12.15$, $g_{a}=83.84$, are determined from the normalization condition of the unpolarized PDFs
\begin{align}
	\pi \int^1_0dx\int^\infty_0d\bm{k}_T^2 {f}_1^{q(X)}(x,\bm k_T^2)=1.
\end{align}

\subsection{Twist-3 GPDs}
\begin{figure*}
	\centering
	\includegraphics[width=0.40\columnwidth]{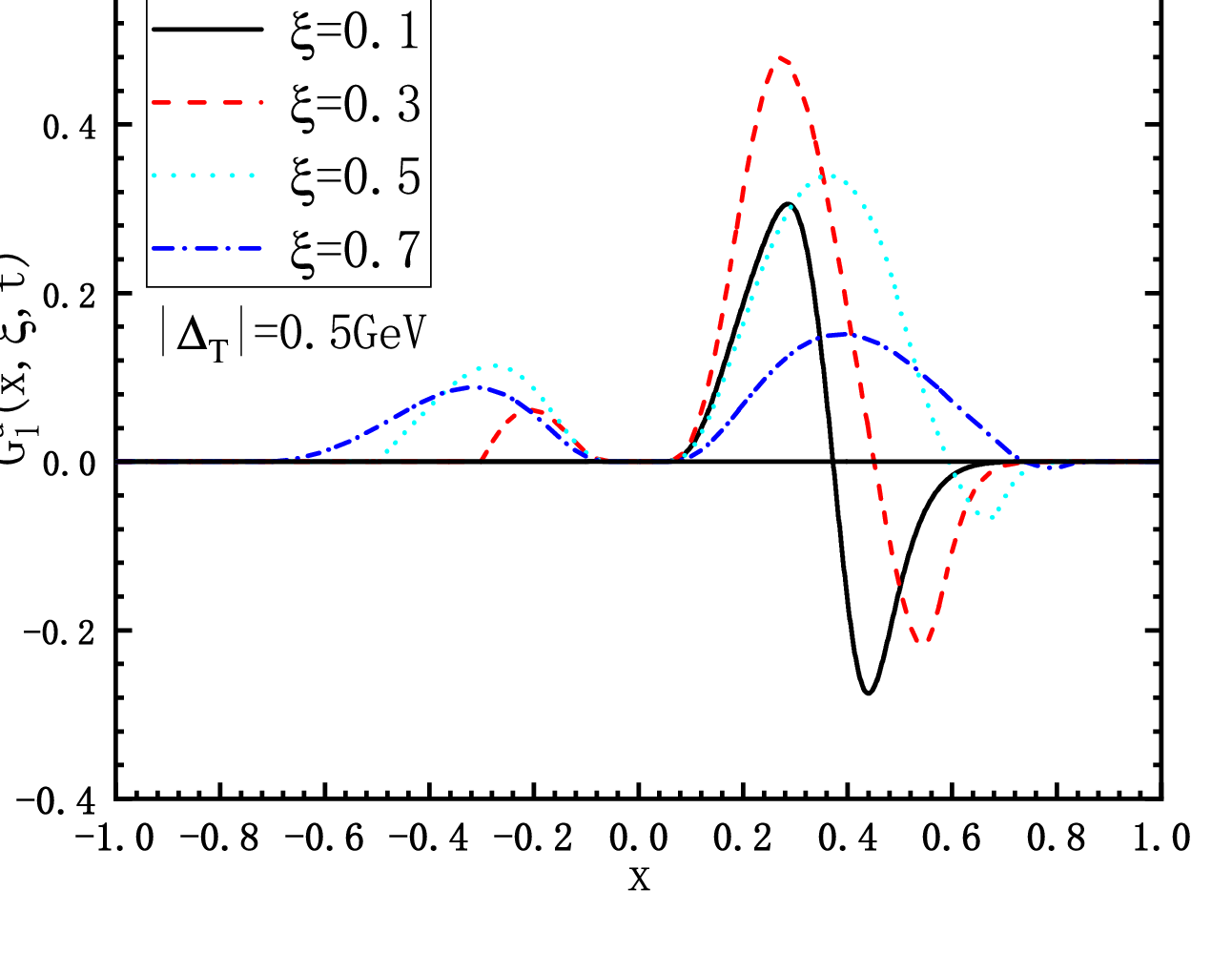} 
	\includegraphics[width=0.40\columnwidth]{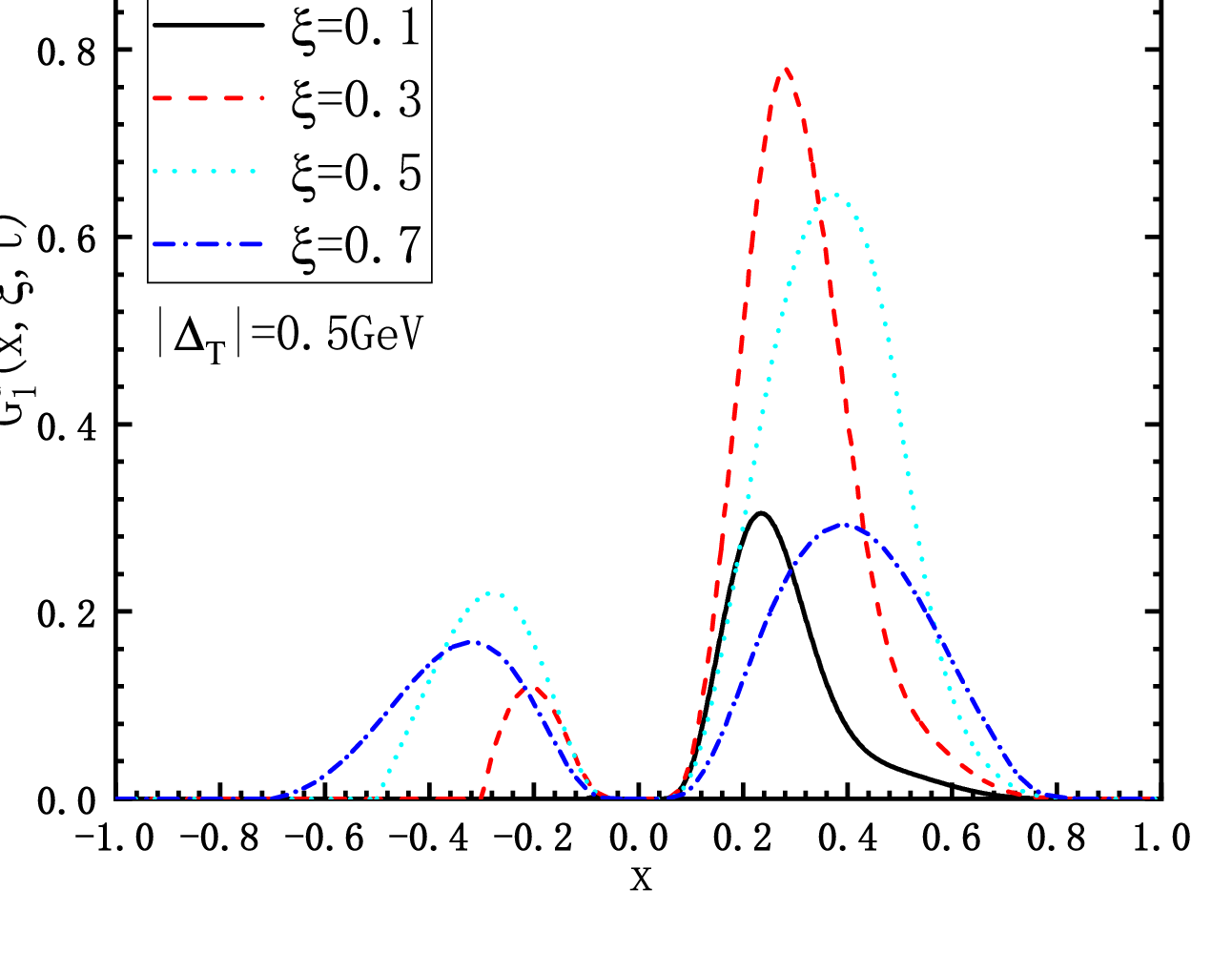}\\
	\includegraphics[width=0.40\columnwidth]{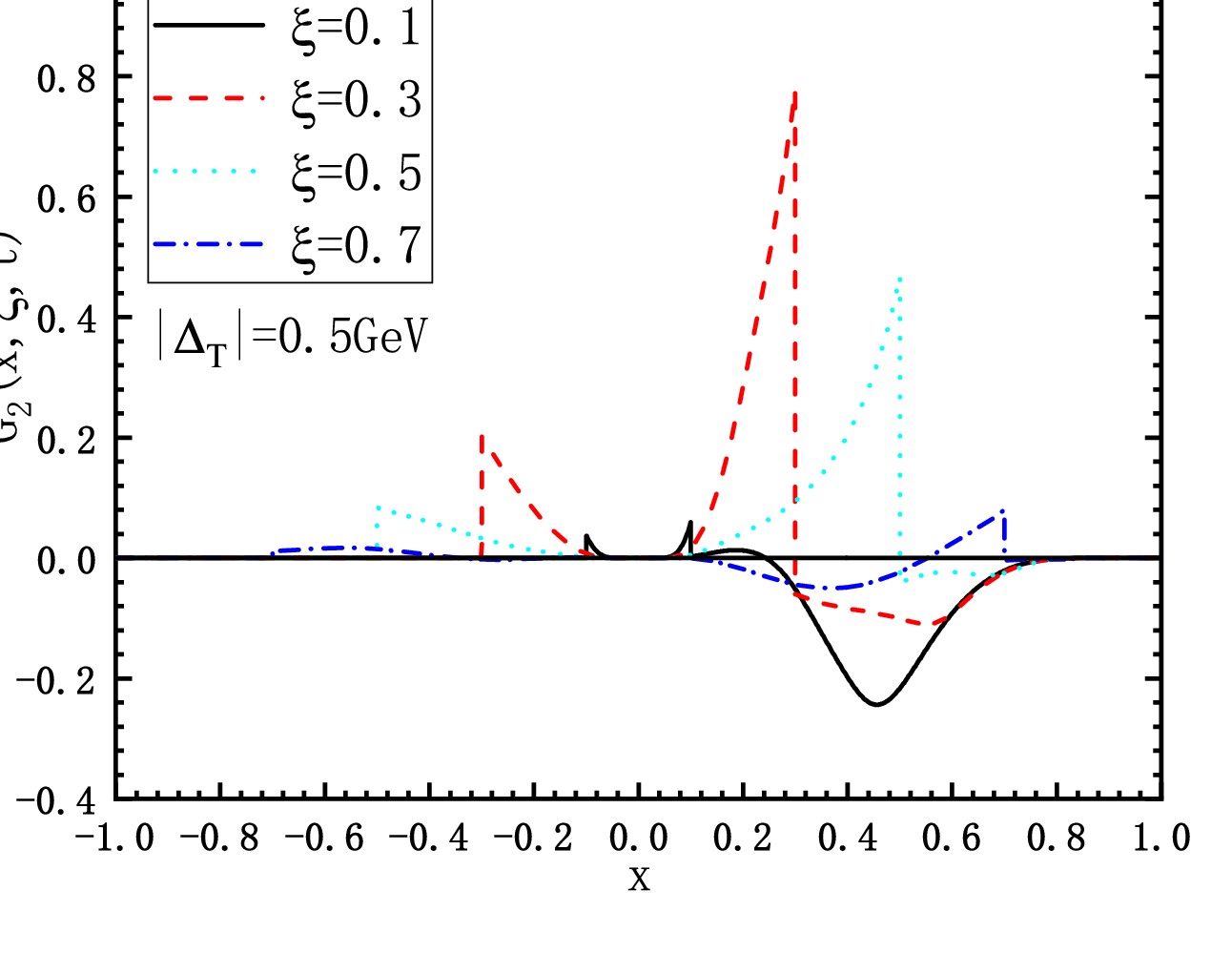}
	\includegraphics[width=0.40\columnwidth]{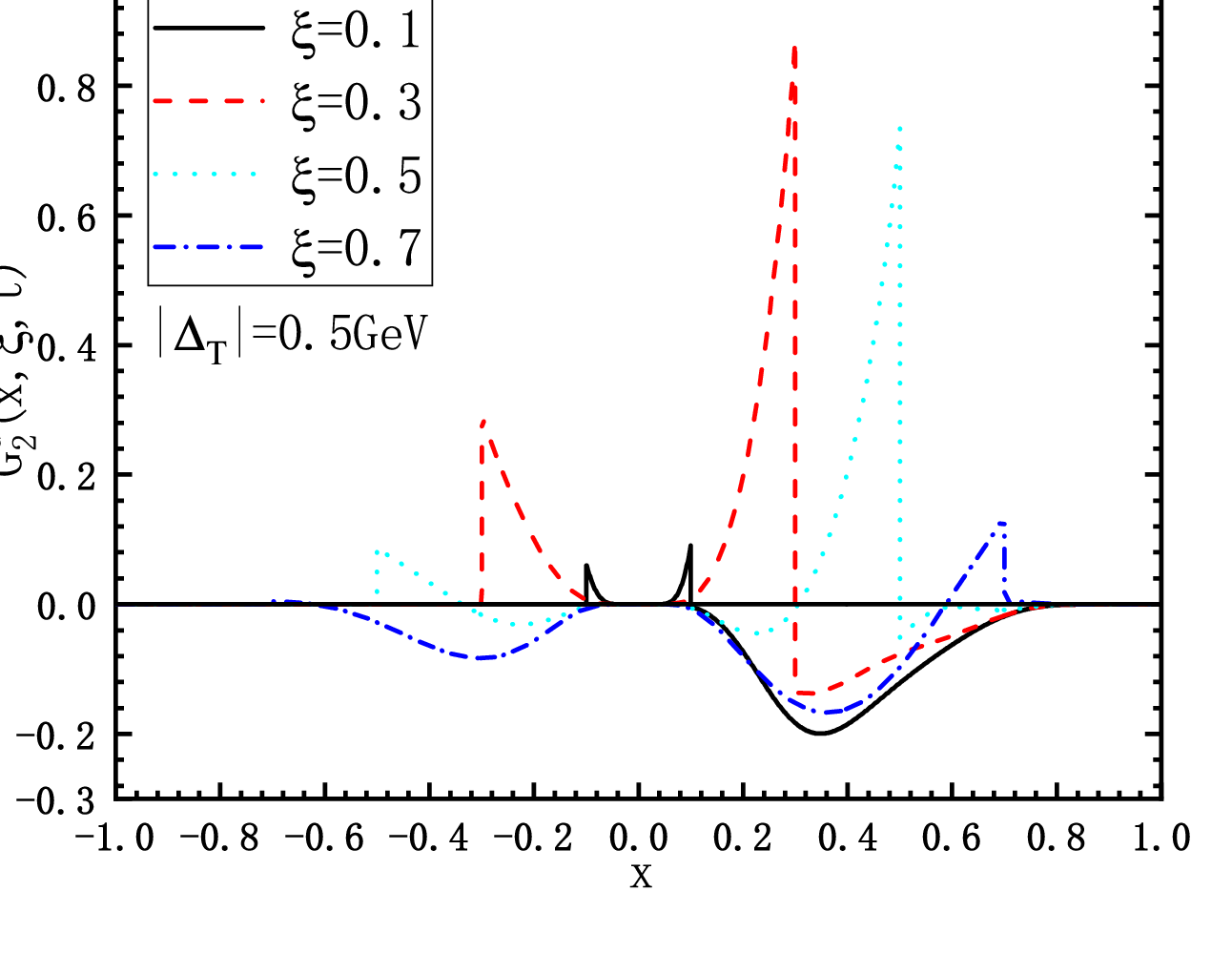}\\
	\includegraphics[width=0.40\columnwidth]{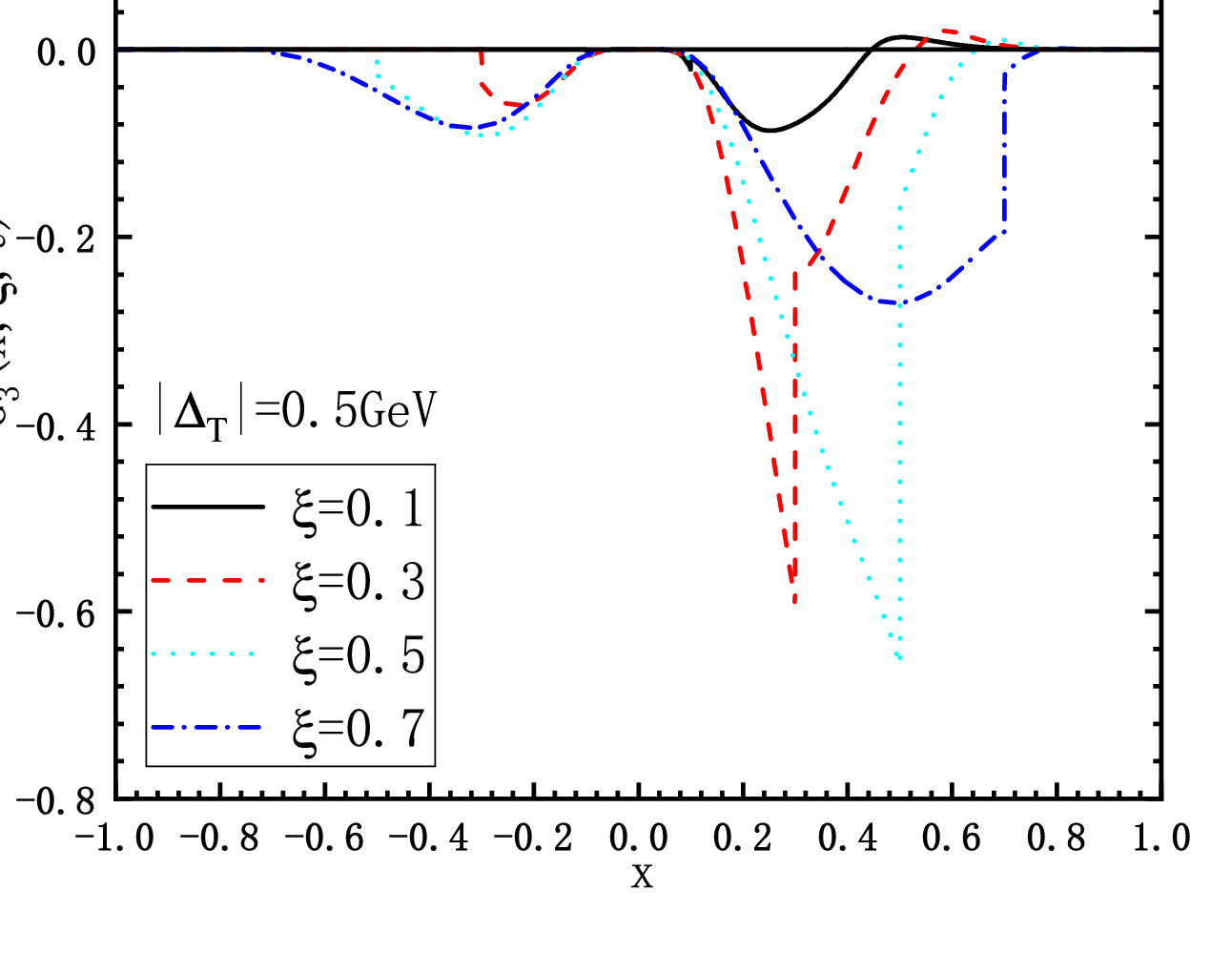}
	\includegraphics[width=0.40\columnwidth]{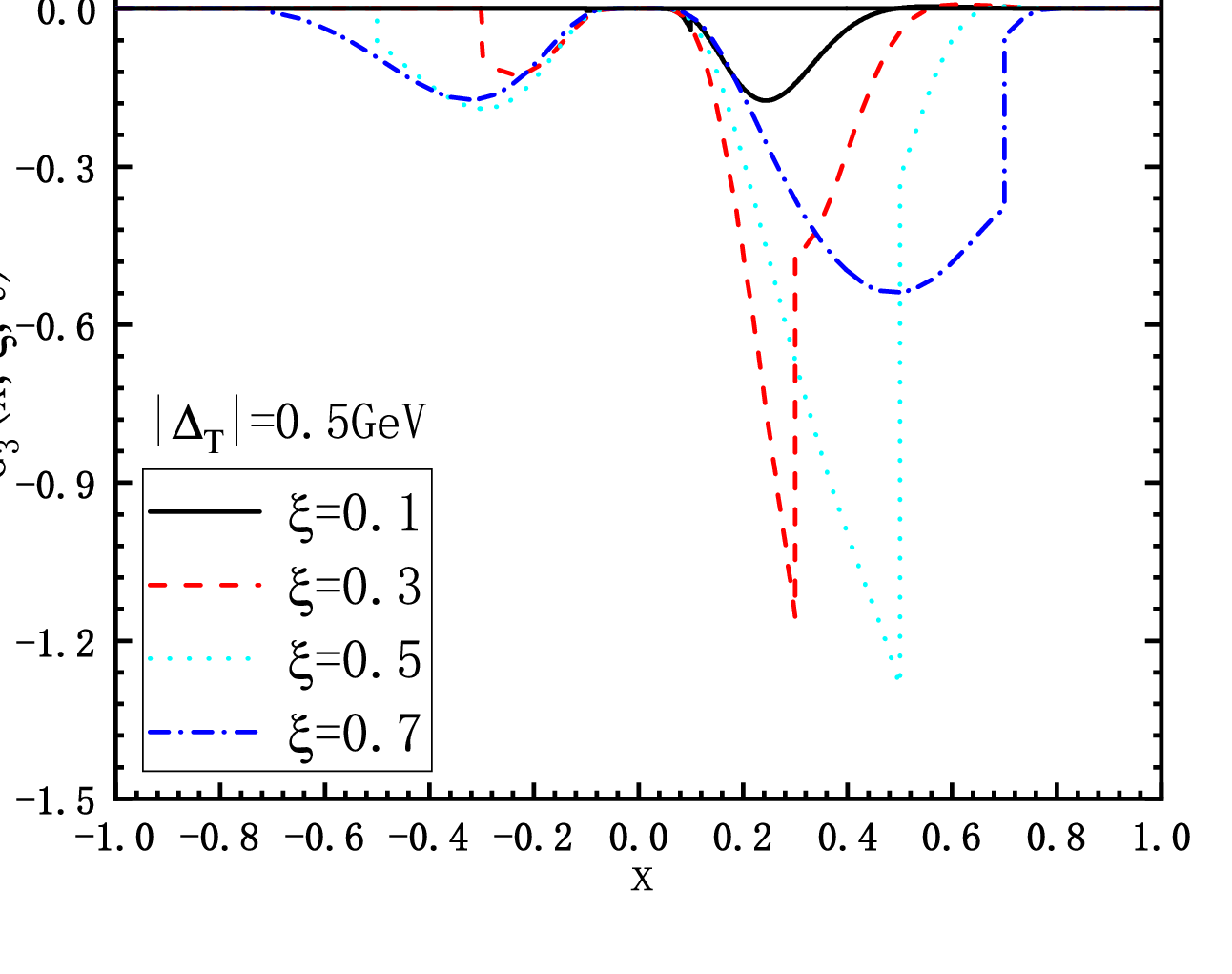}\\
	\includegraphics[width=0.40\columnwidth]{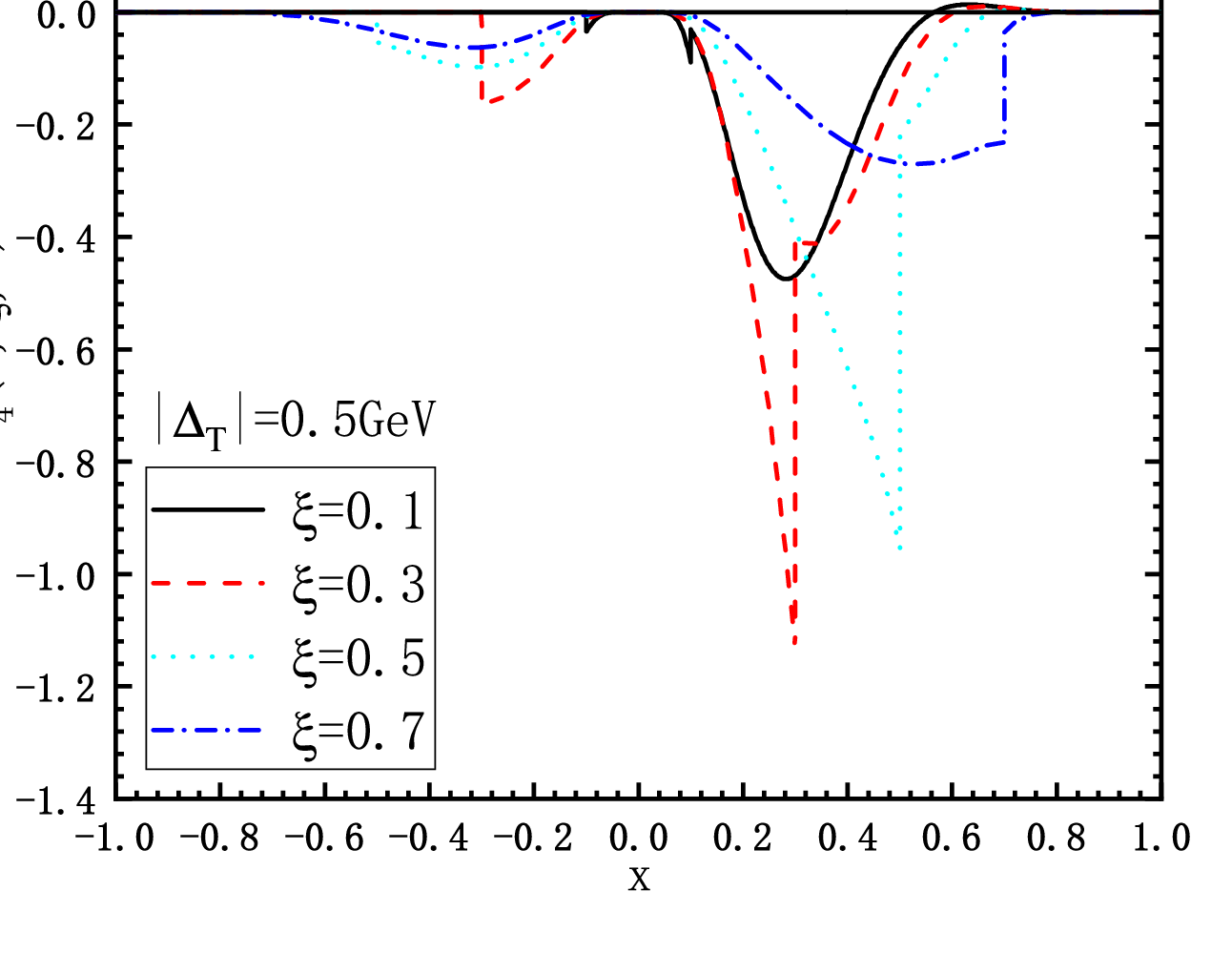}
	\includegraphics[width=0.40\columnwidth]{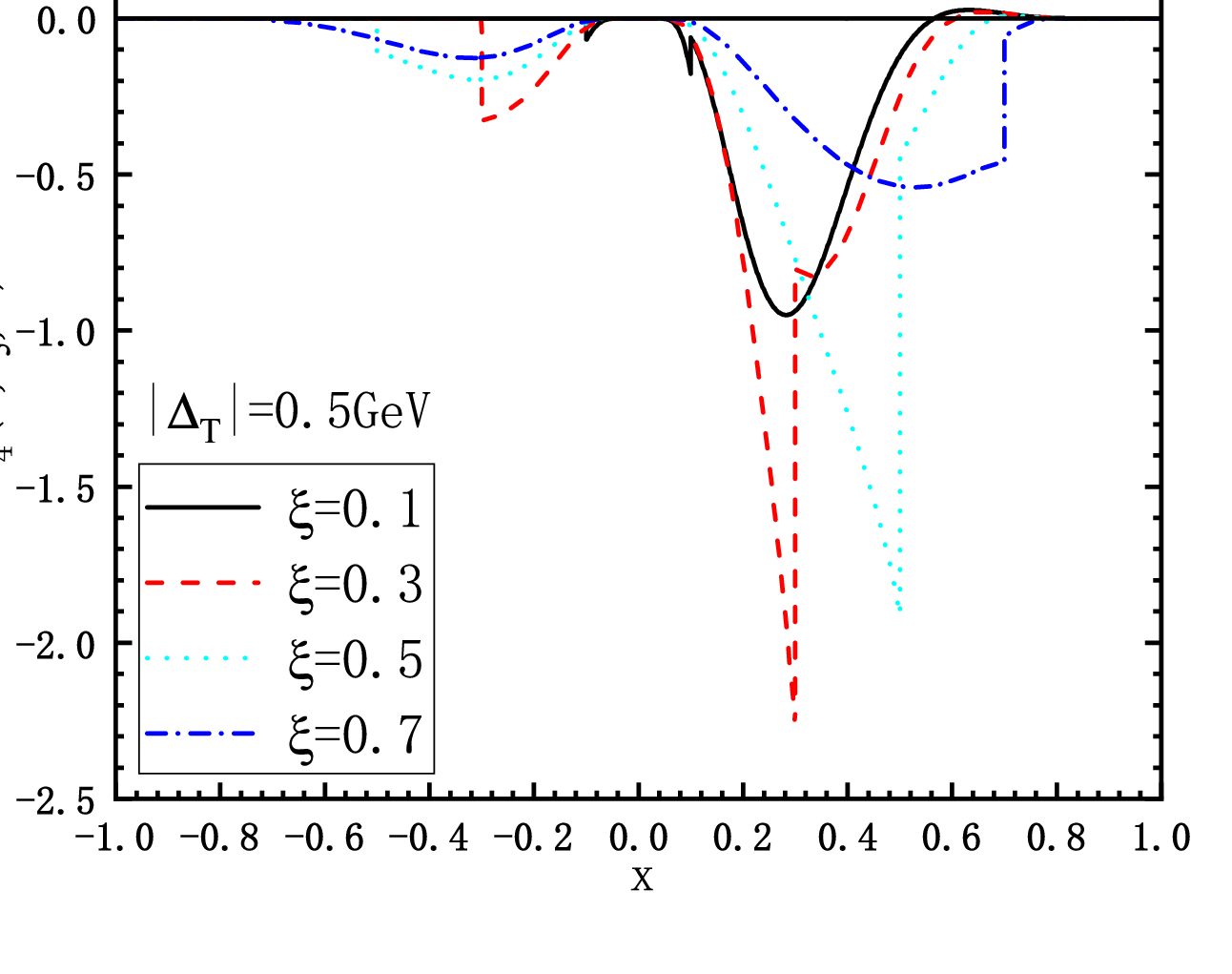}\\
	\caption{The vector twist-3 GPDs $G_1(x,\xi,t)$, $G_2(x,\xi,t)$, $G_3(x,\xi,t)$ and $G_4(x,\xi,t)$ of the $u$ (left panel) and $d$ (right panel) quarks as functions of $x$ at fixed $|\bm{\Delta}_T|=0.5~\text{GeV}$ for $\xi=0.1$, 0.3, 0.5, 0.7, respectively.}
	\label{fig:vectorGPDx}
\end{figure*}
\begin{figure}
	\centering
	\includegraphics[width=0.40\columnwidth]{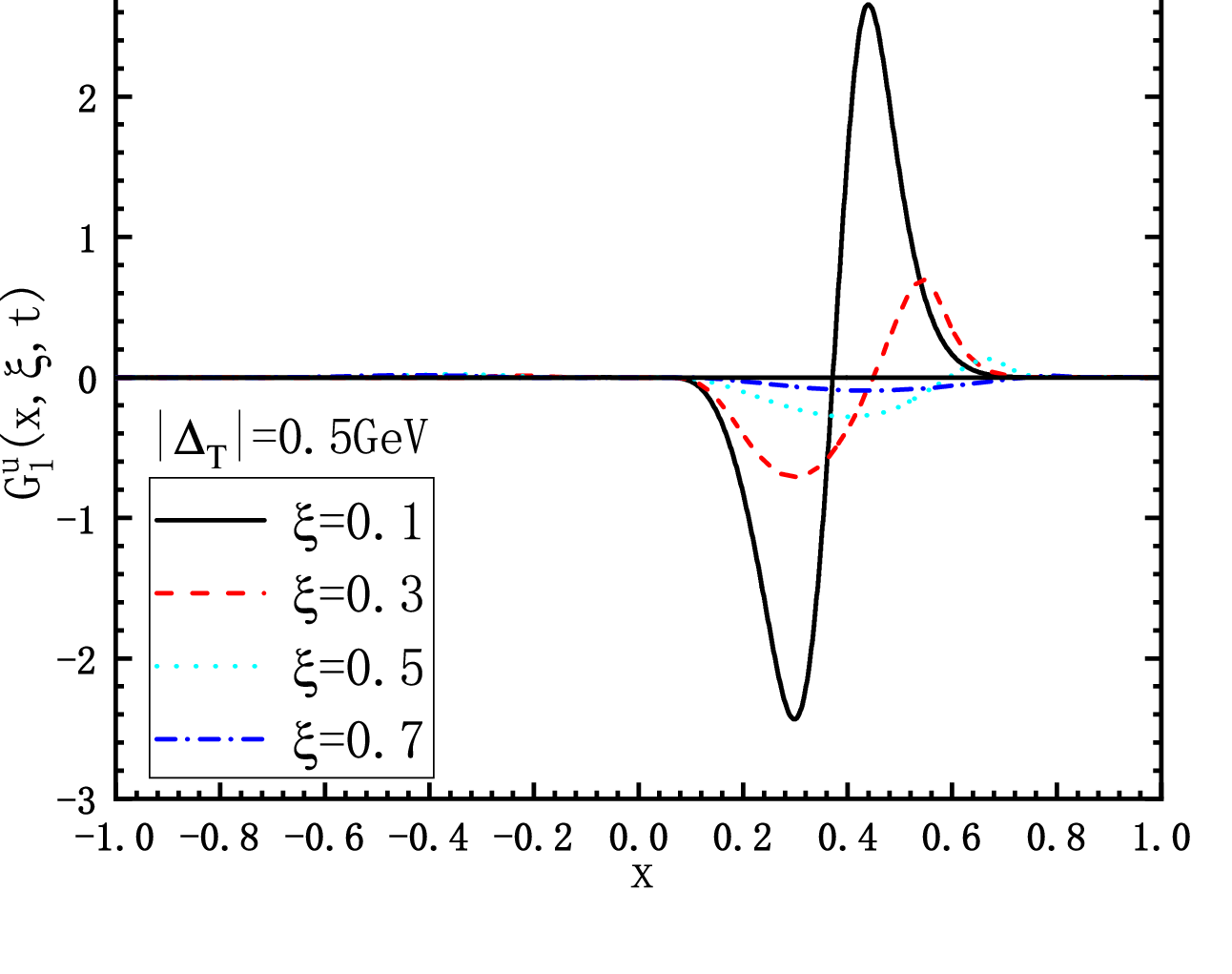}
	\includegraphics[width=0.40\columnwidth]{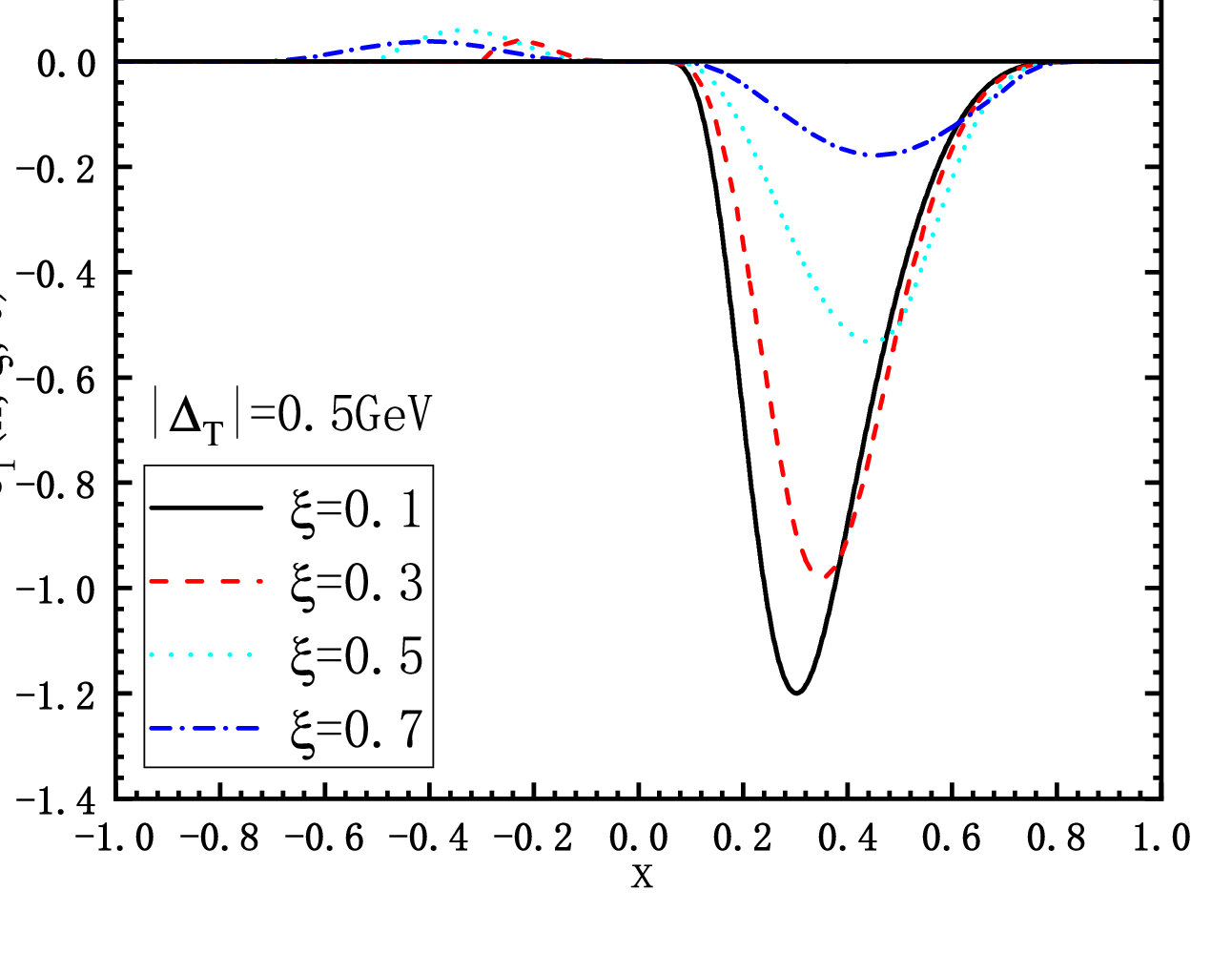}\\
	\includegraphics[width=0.40\columnwidth]{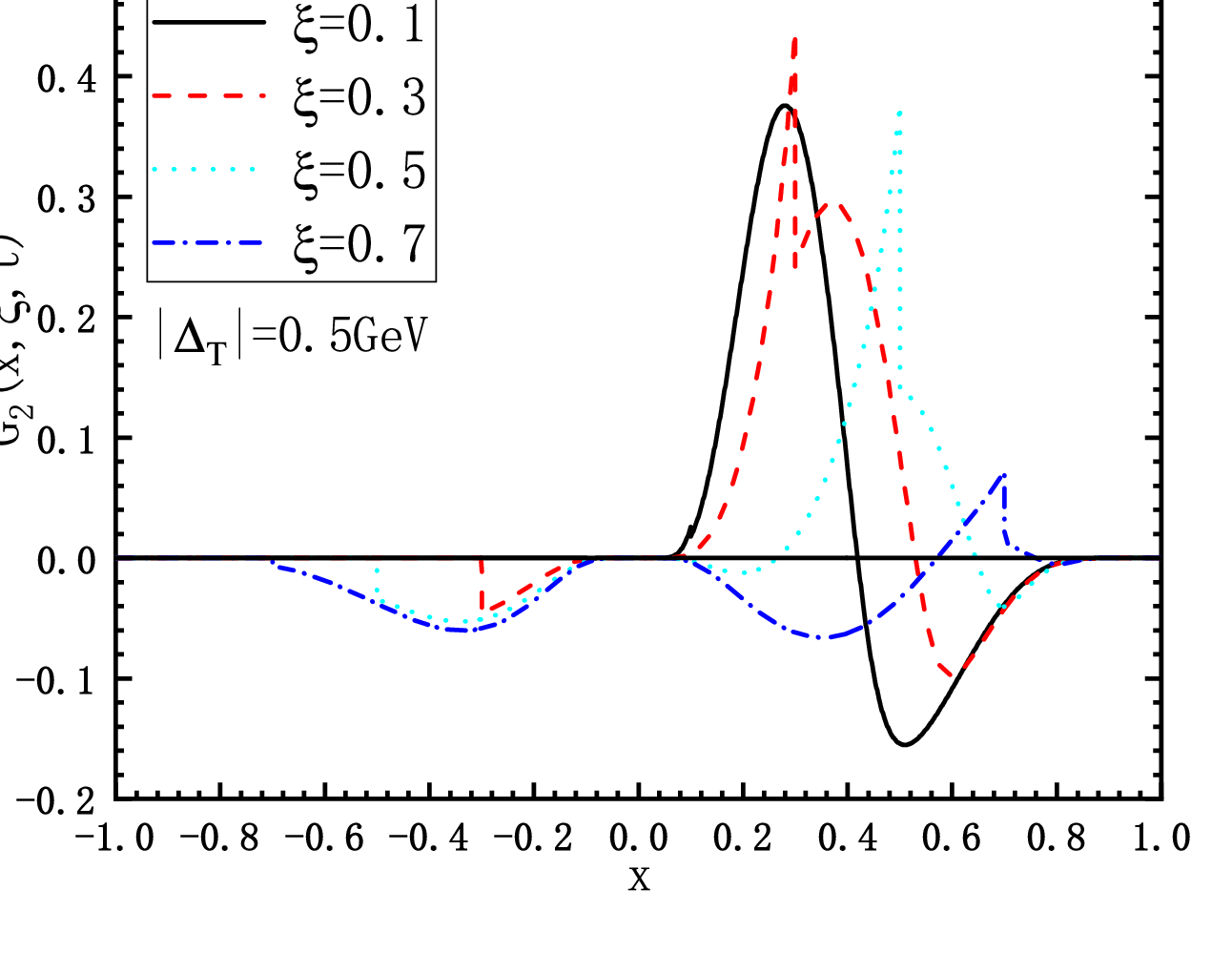}
	\includegraphics[width=0.40\columnwidth]{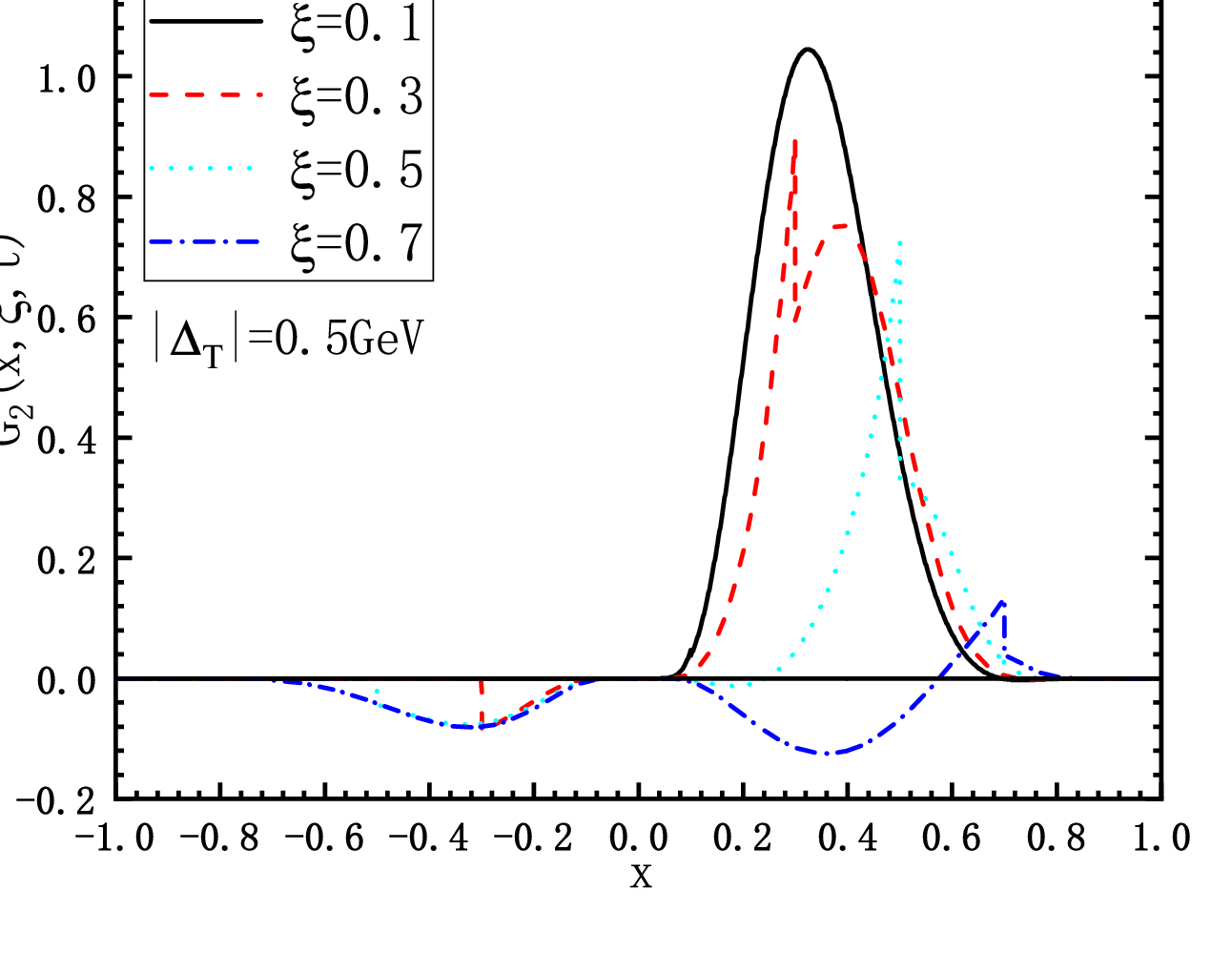}\\
	\includegraphics[width=0.40\columnwidth]{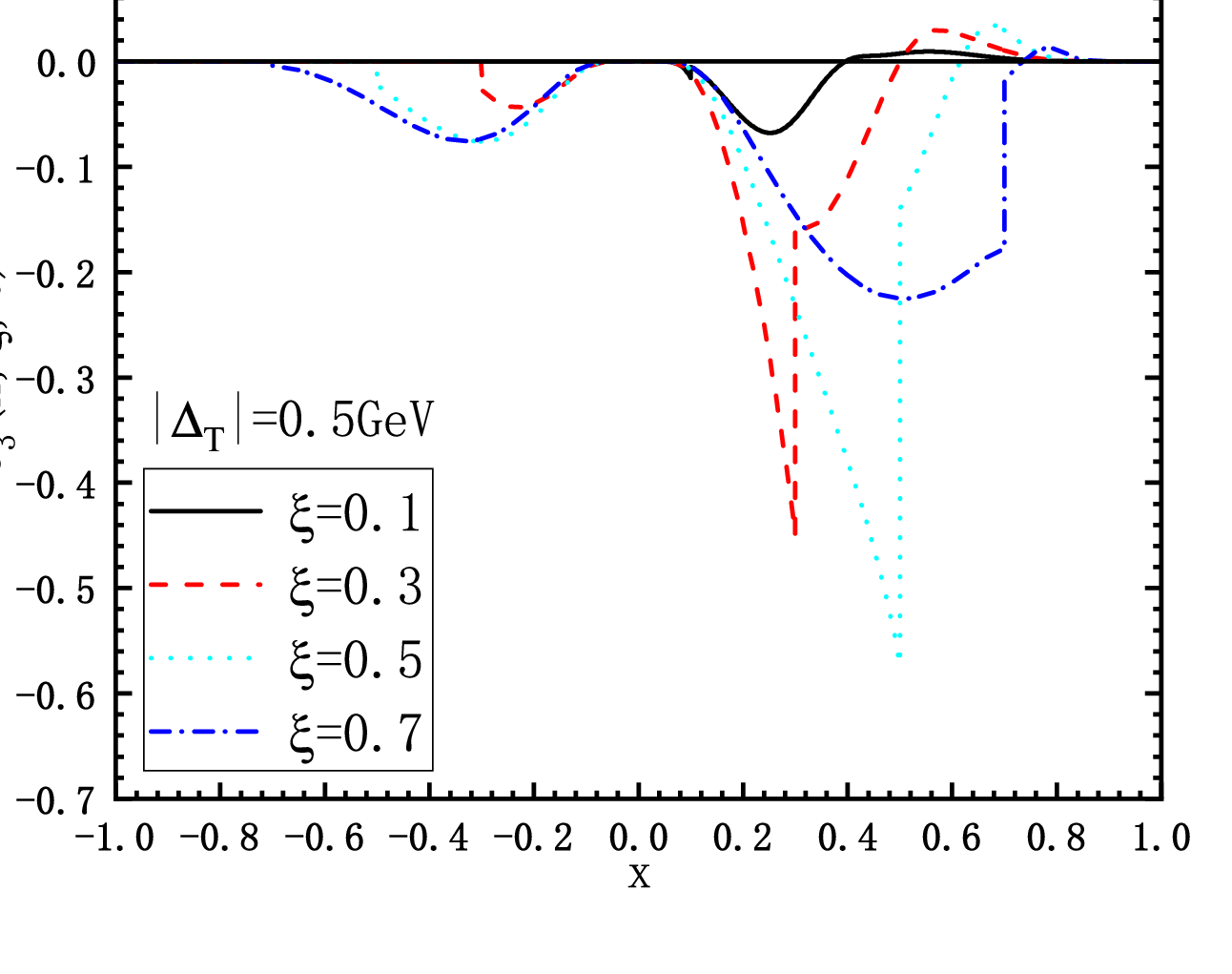}
	\includegraphics[width=0.40\columnwidth]{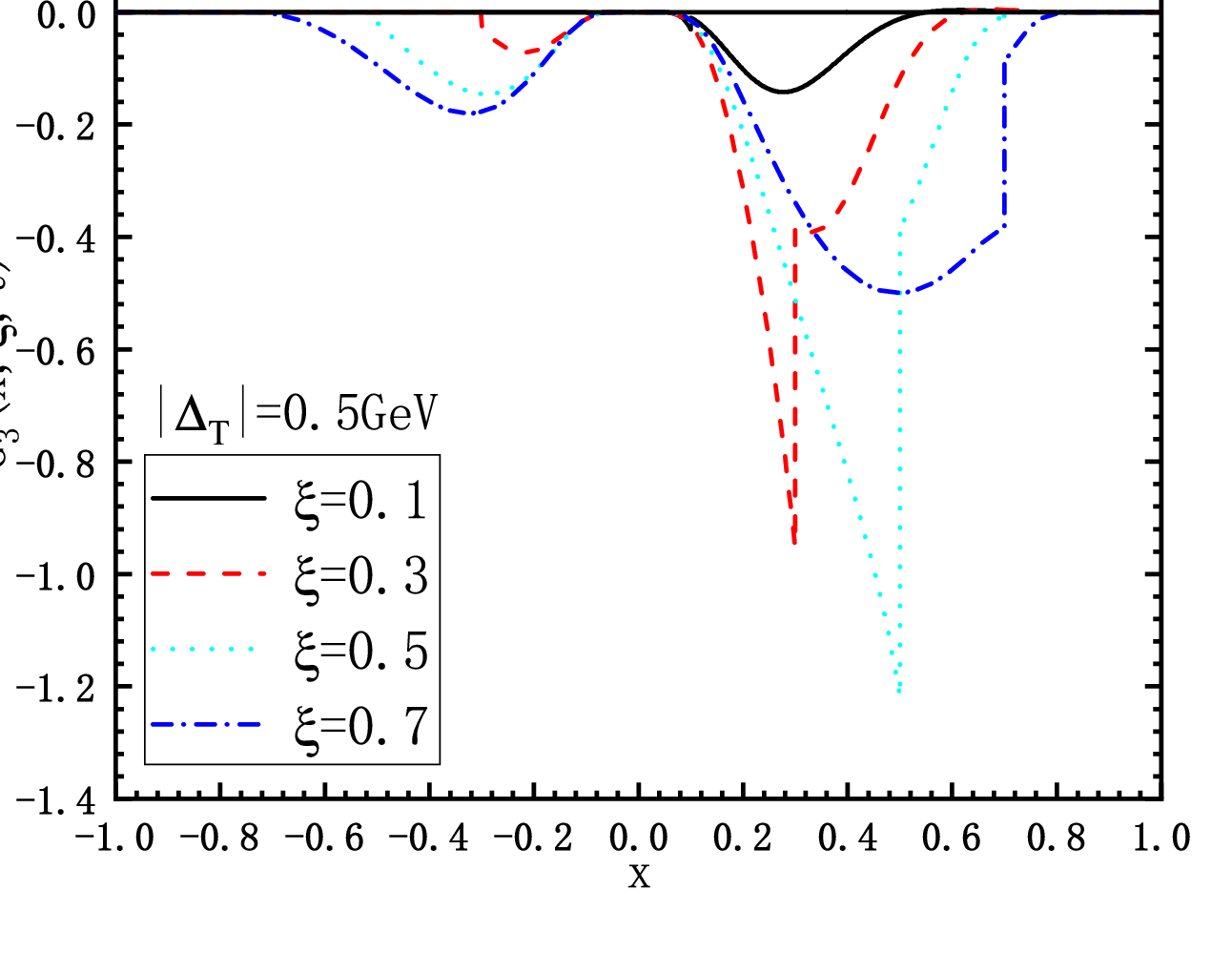}\\
	\includegraphics[width=0.40\columnwidth]{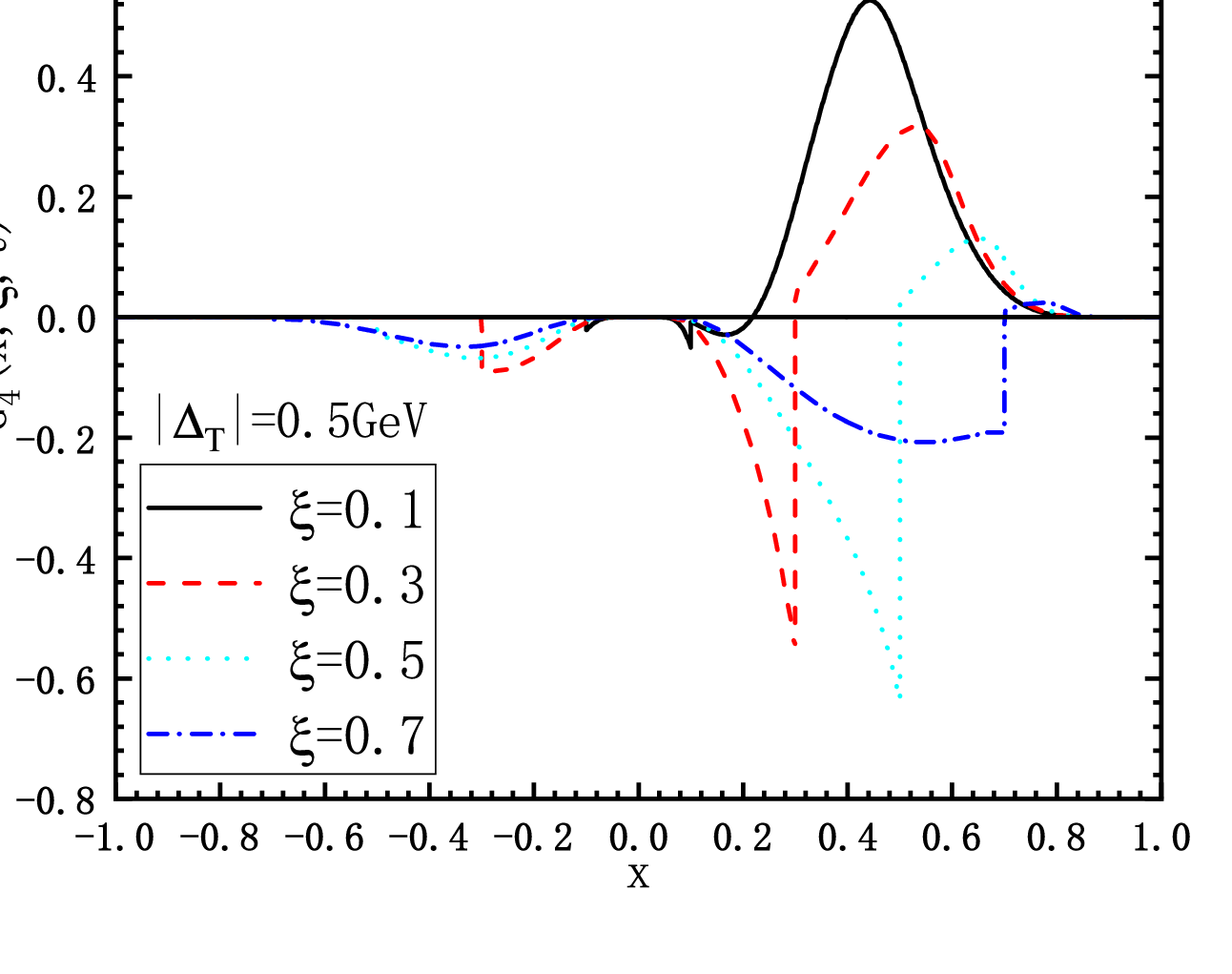}
	\includegraphics[width=0.40\columnwidth]{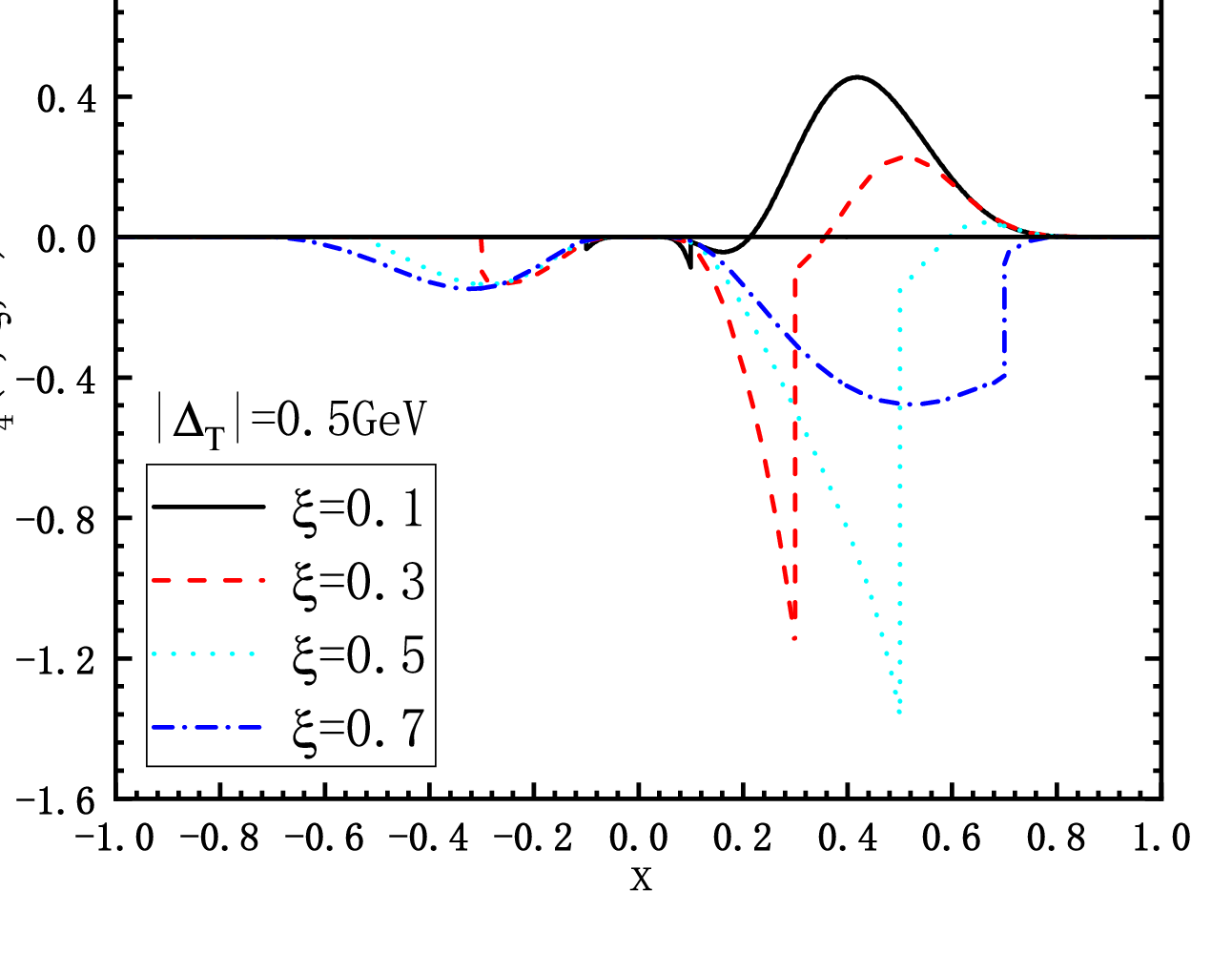}\\
	\caption{The axial-vector twist-3 GPDs $\tilde{G}_1(x,\xi,t)$, $\tilde{G}_2(x,\xi,t)$, $\tilde{G}_3(x,\xi,t)$ and $\tilde{G}_4(x,\xi,t)$ of the $u$ (left panel) and $d$ (right panel) quarks as functions of $x$ at fixed $|\bm{\Delta}_T|=0.5~\text{GeV}$ for $\xi=0.1$, 0.3, 0.5, 0.7, respectively.}
	\label{fig:axialvectorGPDx}
\end{figure}

\begin{figure}
	\centering
	\includegraphics[width=0.40\columnwidth]{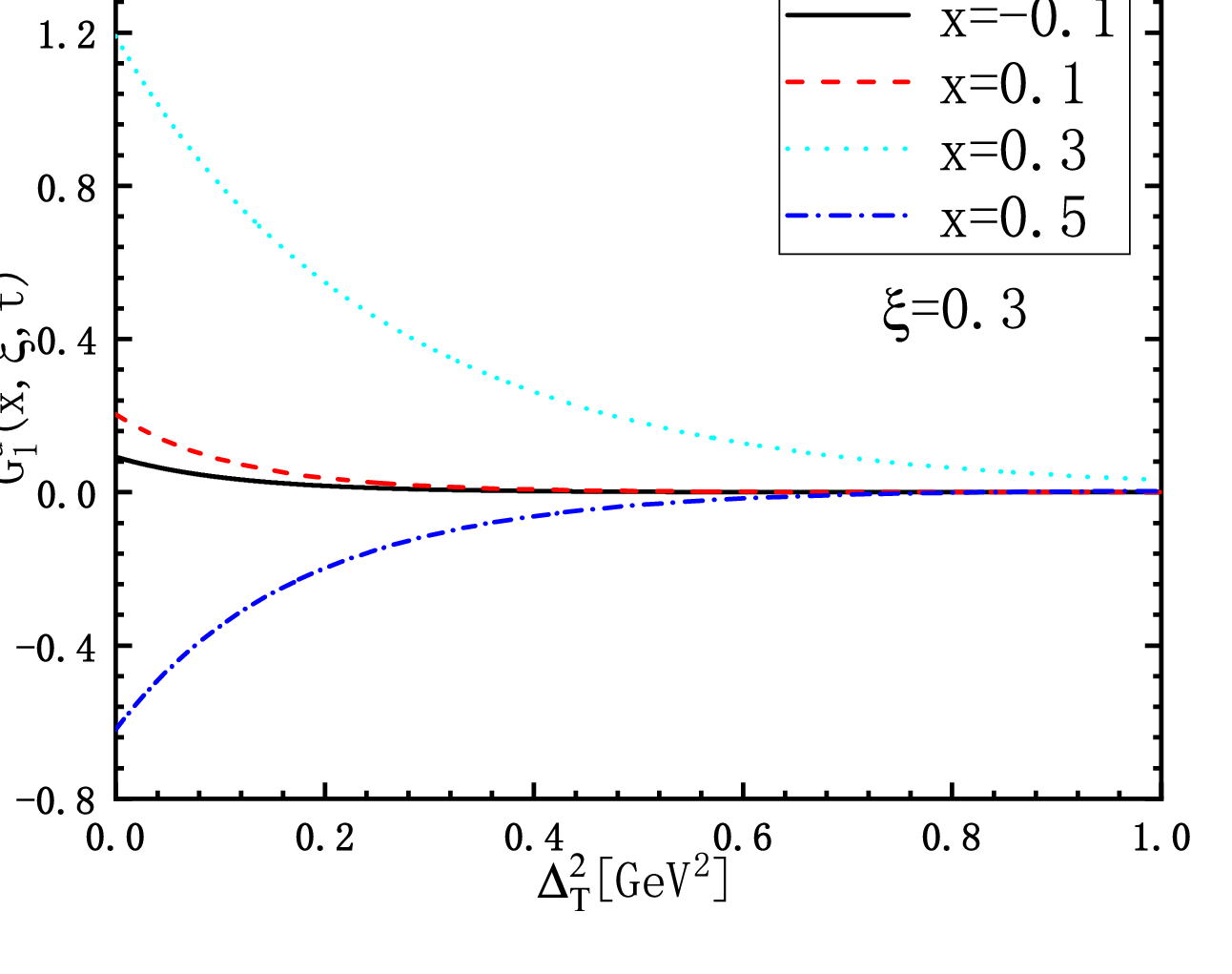}
	\includegraphics[width=0.40\columnwidth]{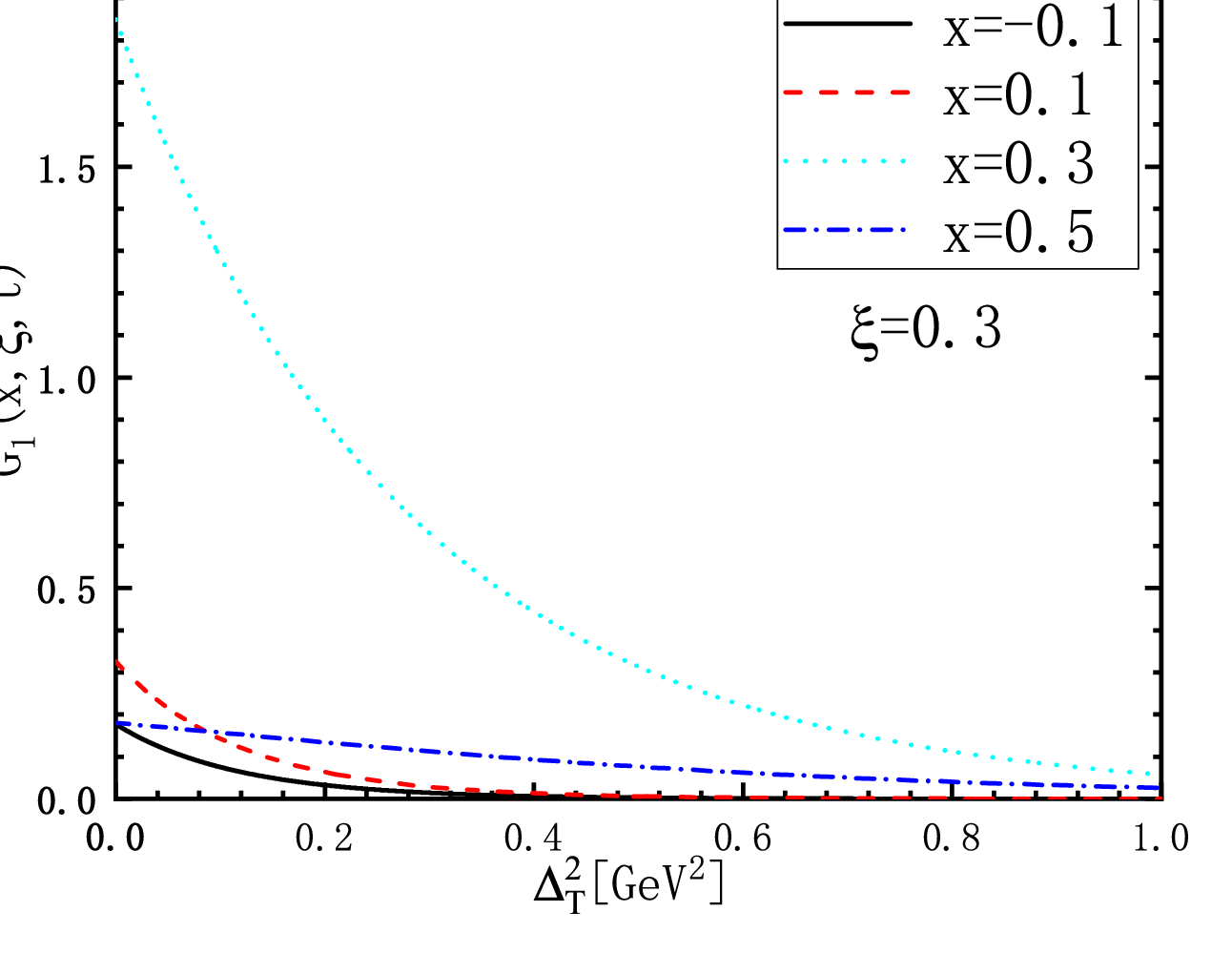}\\
	\includegraphics[width=0.40\columnwidth]{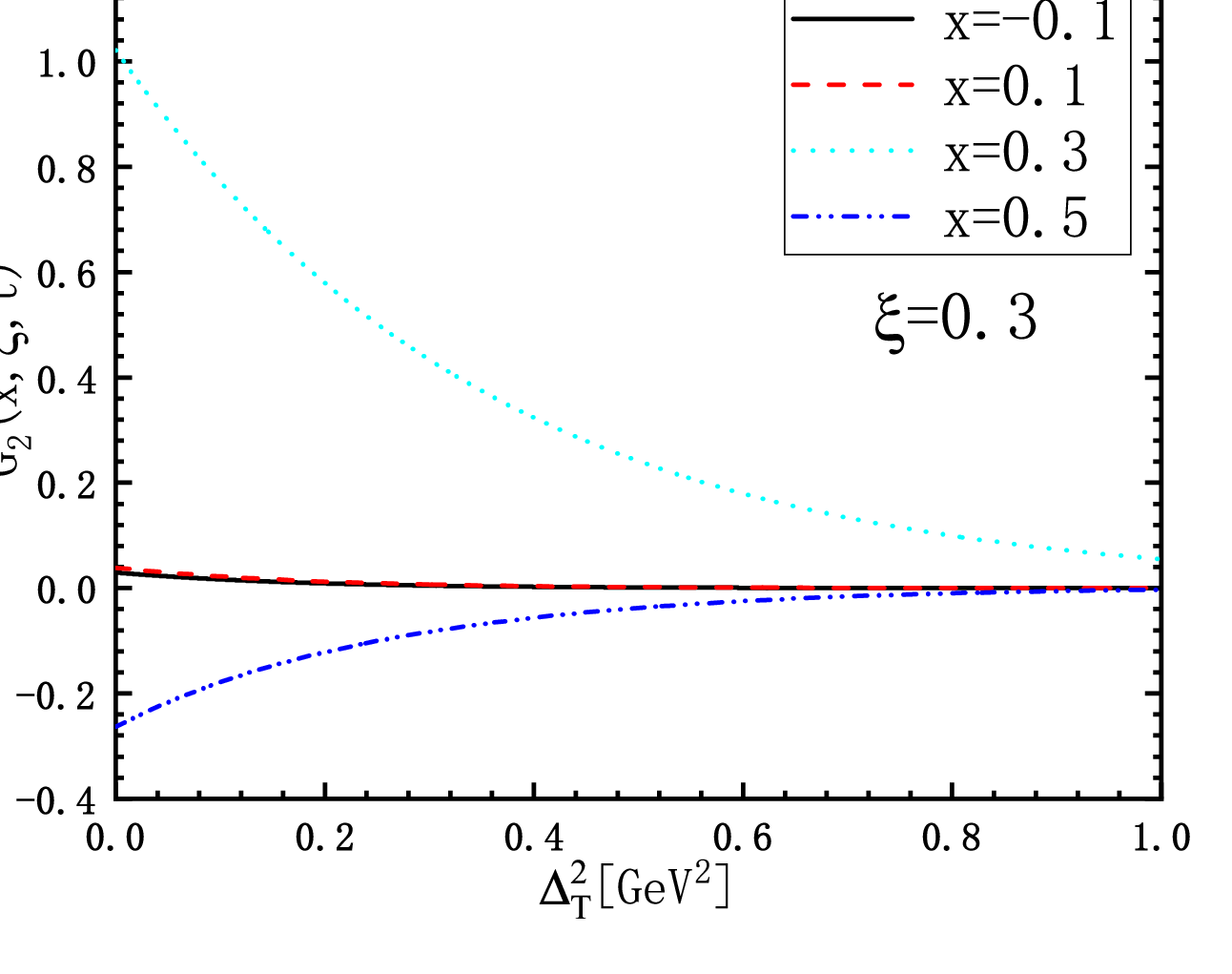}
	\includegraphics[width=0.40\columnwidth]{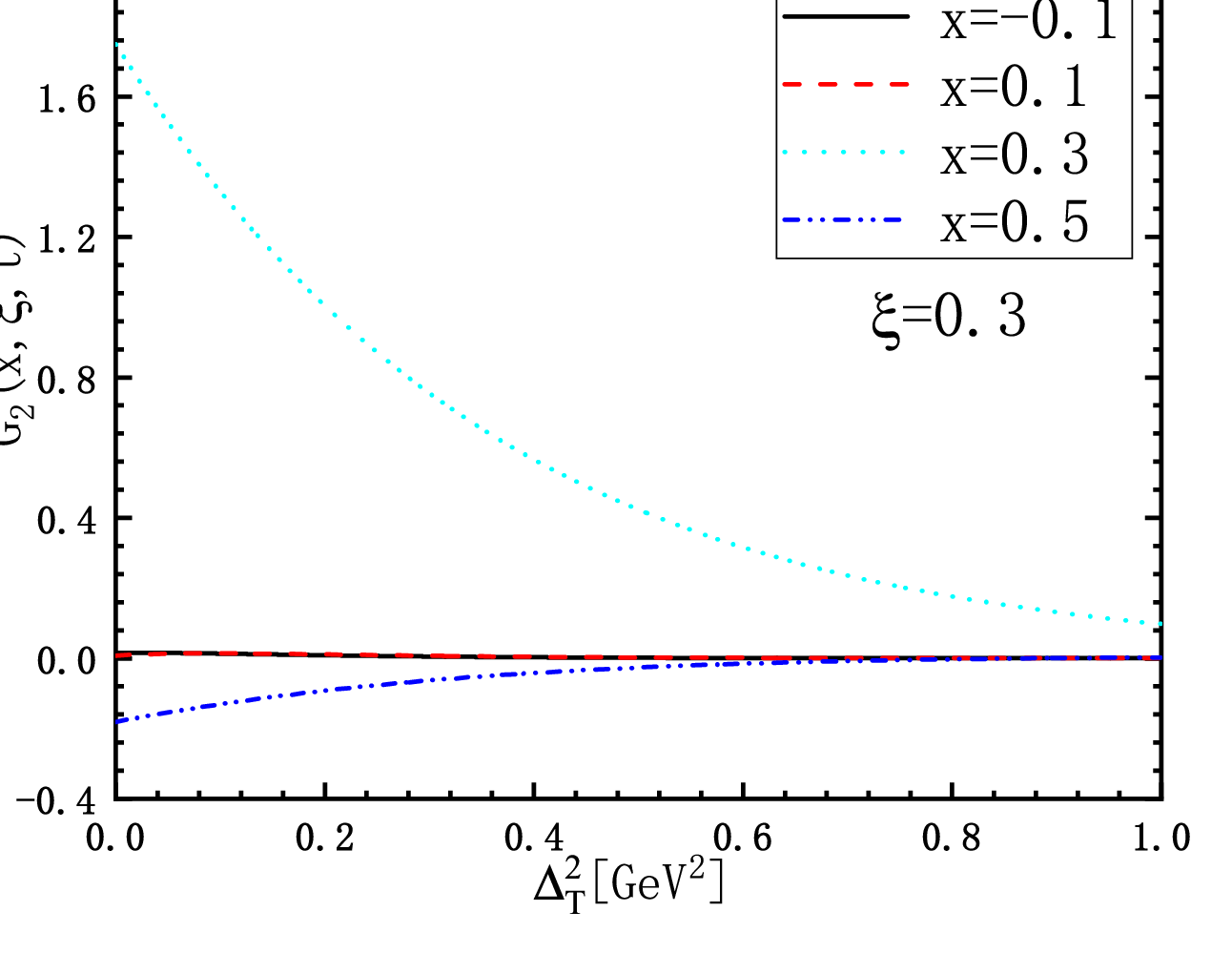}\\
	\includegraphics[width=0.40\columnwidth]{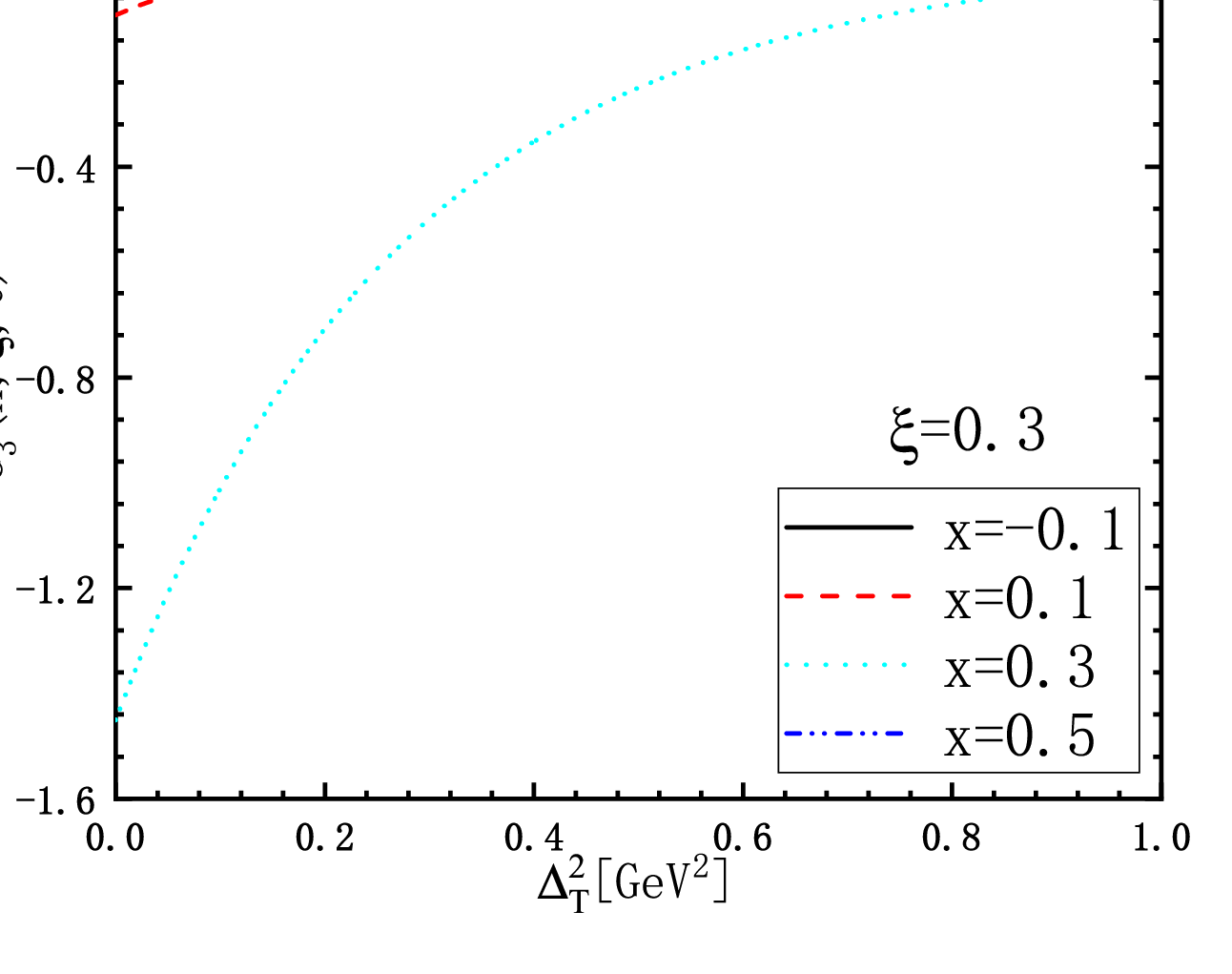}
	\includegraphics[width=0.40\columnwidth]{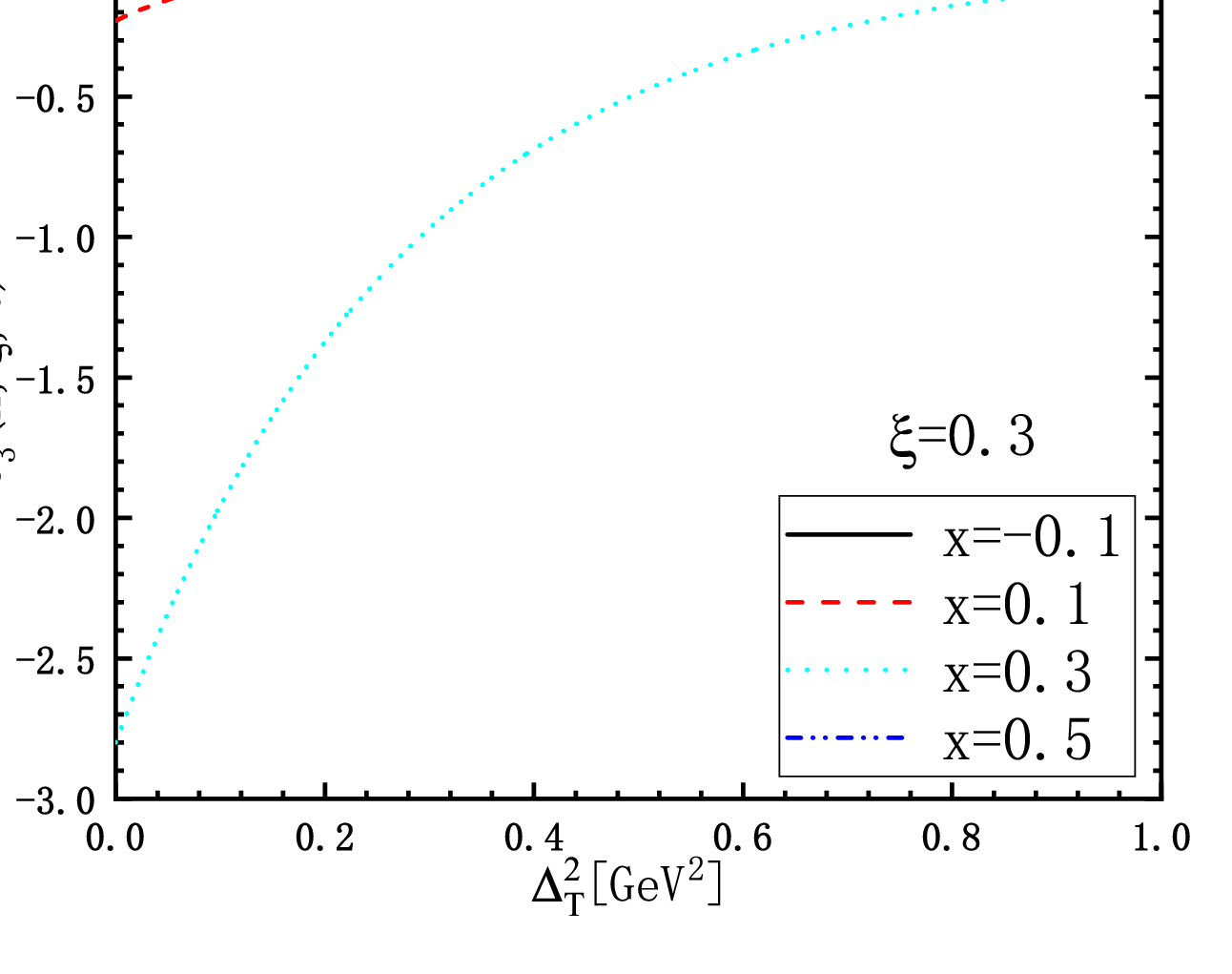}\\
	\includegraphics[width=0.40\columnwidth]{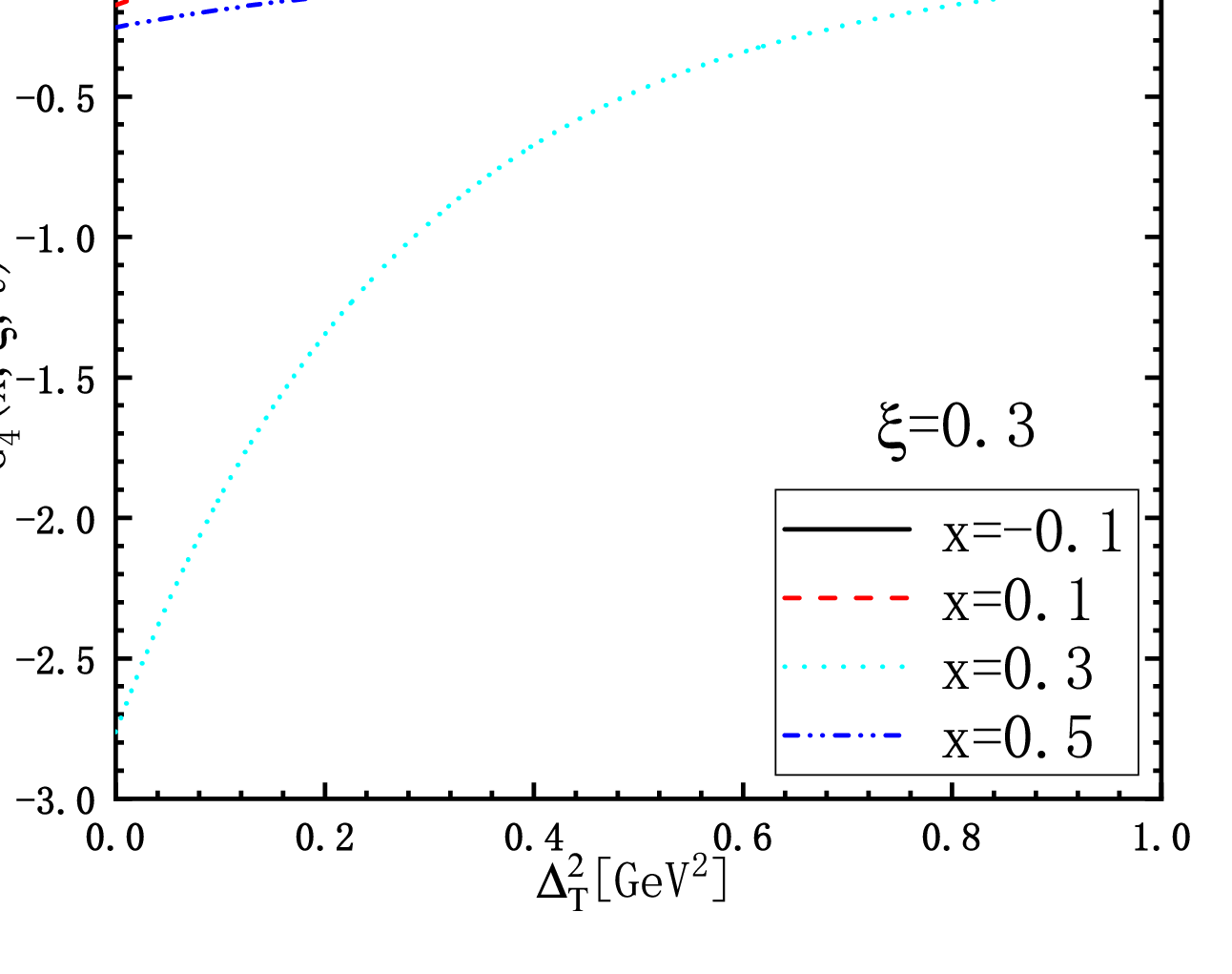}
	\includegraphics[width=0.40\columnwidth]{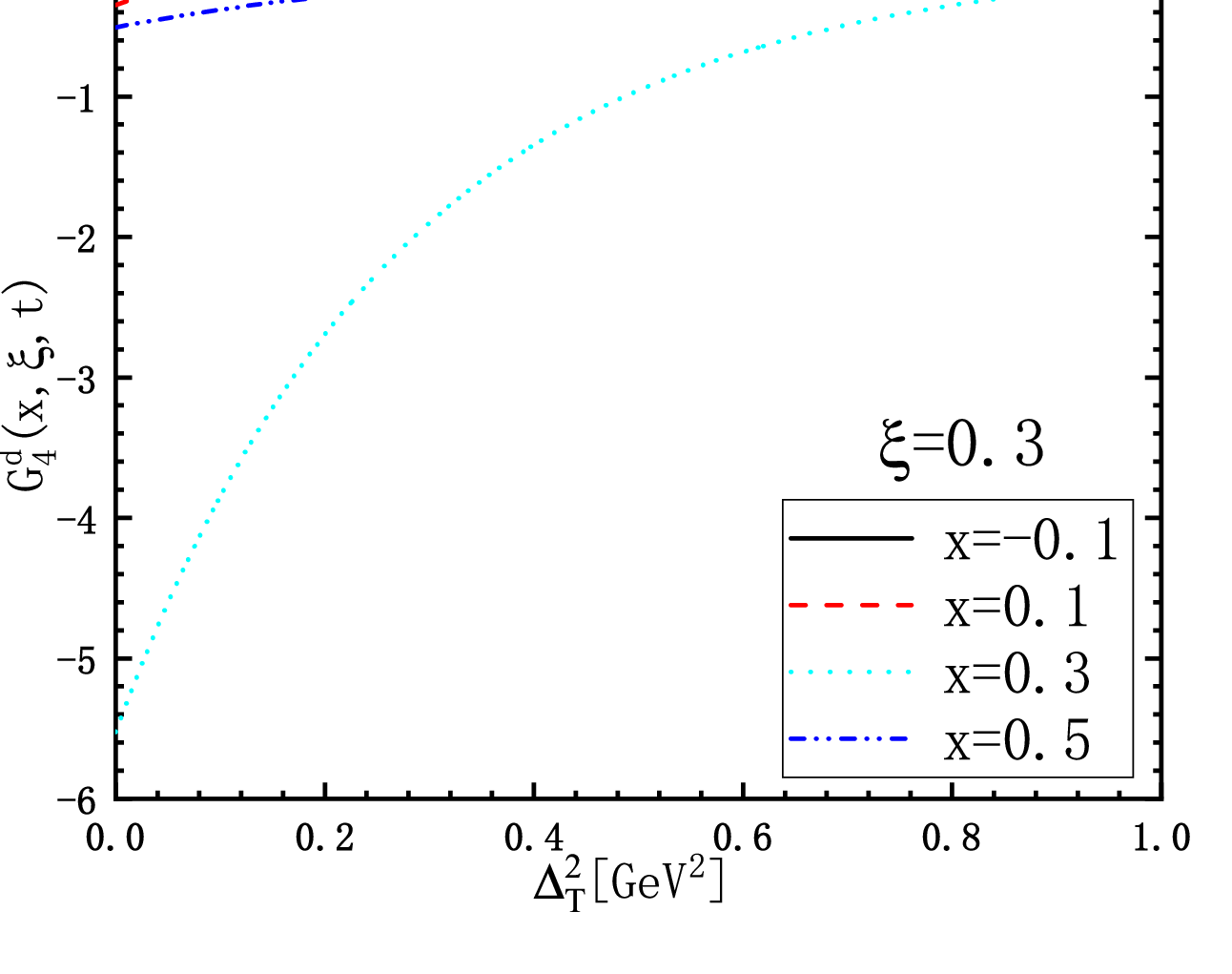}\\
	\caption{The vector twist-3 GPDs $G_1(x,\xi,t)$, $G_2(x,\xi,t)$, $G_3(x,\xi,t)$ and $G_4(x,\xi,t)$ of the $u$ (left panel) and $d$ (right panel) quarks as functions of $\bm{\Delta}_T^2$ at fixed $\xi=0.3$ for $x=-0.1$, 0.1, 0.3, 0.5, respectively. }
	\label{fig:vectorGPDdelta}
\end{figure}

\begin{figure}
	\centering
	\includegraphics[width=0.40\columnwidth]{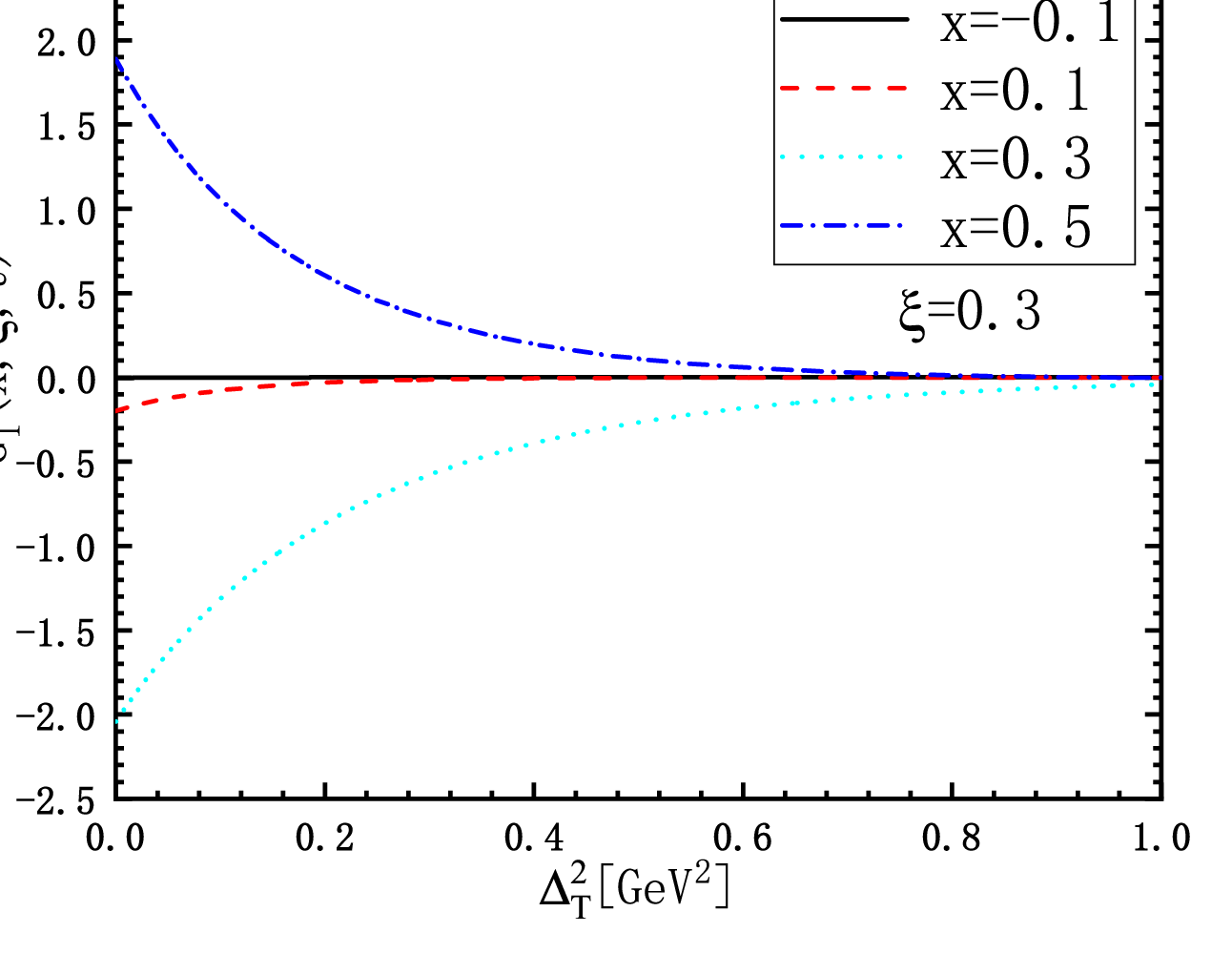}
	\includegraphics[width=0.40\columnwidth]{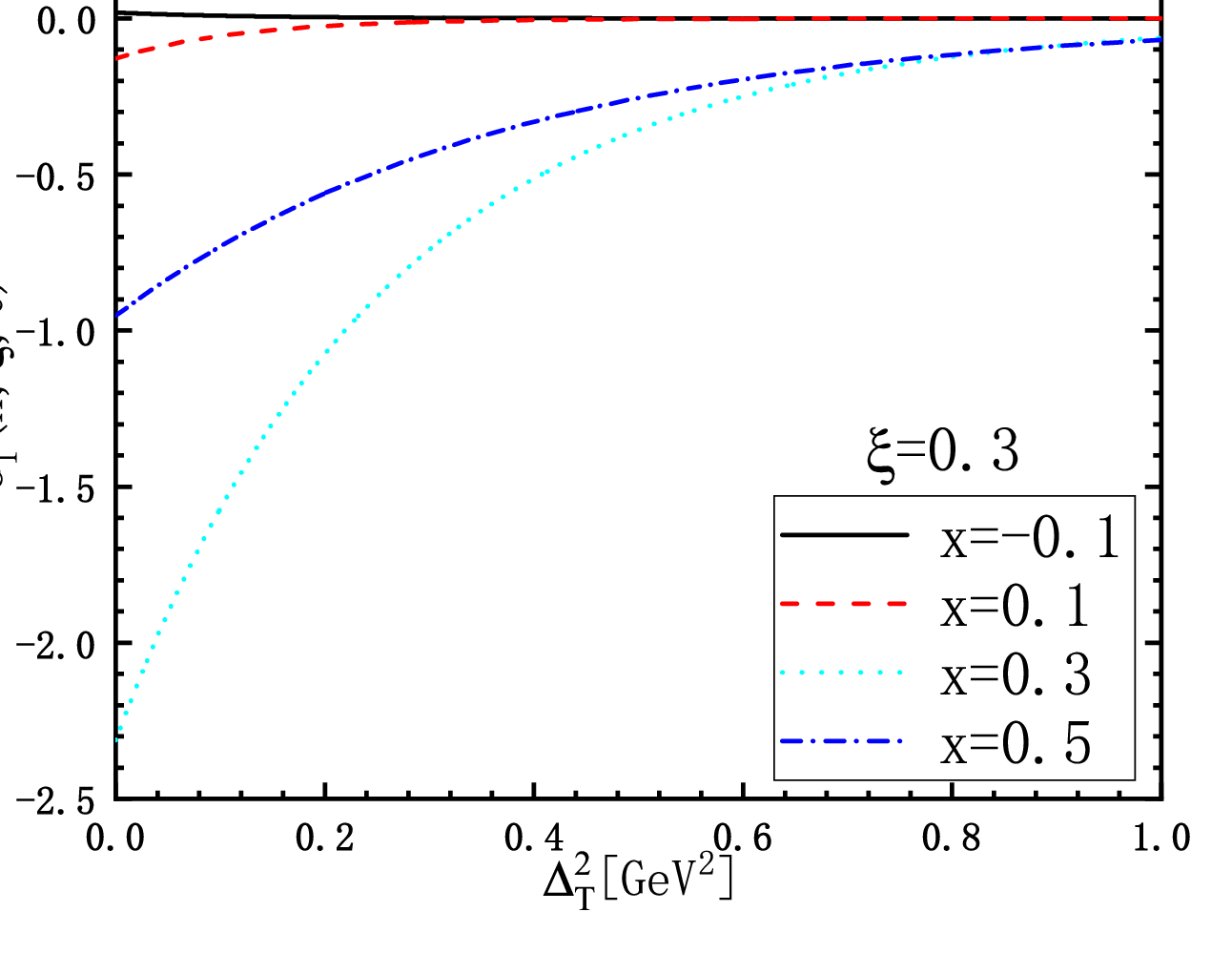}\\
	\includegraphics[width=0.40\columnwidth]{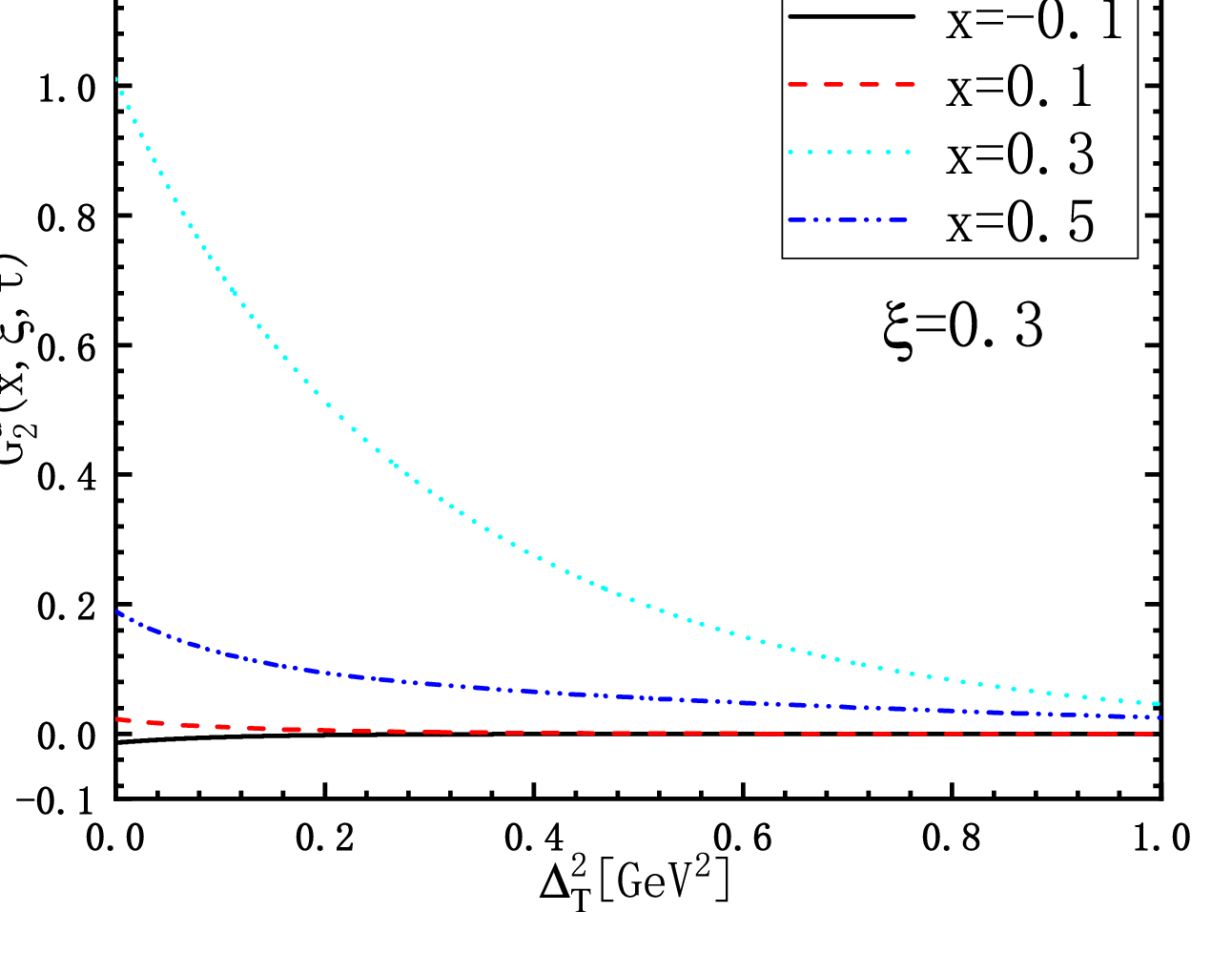}
	\includegraphics[width=0.40\columnwidth]{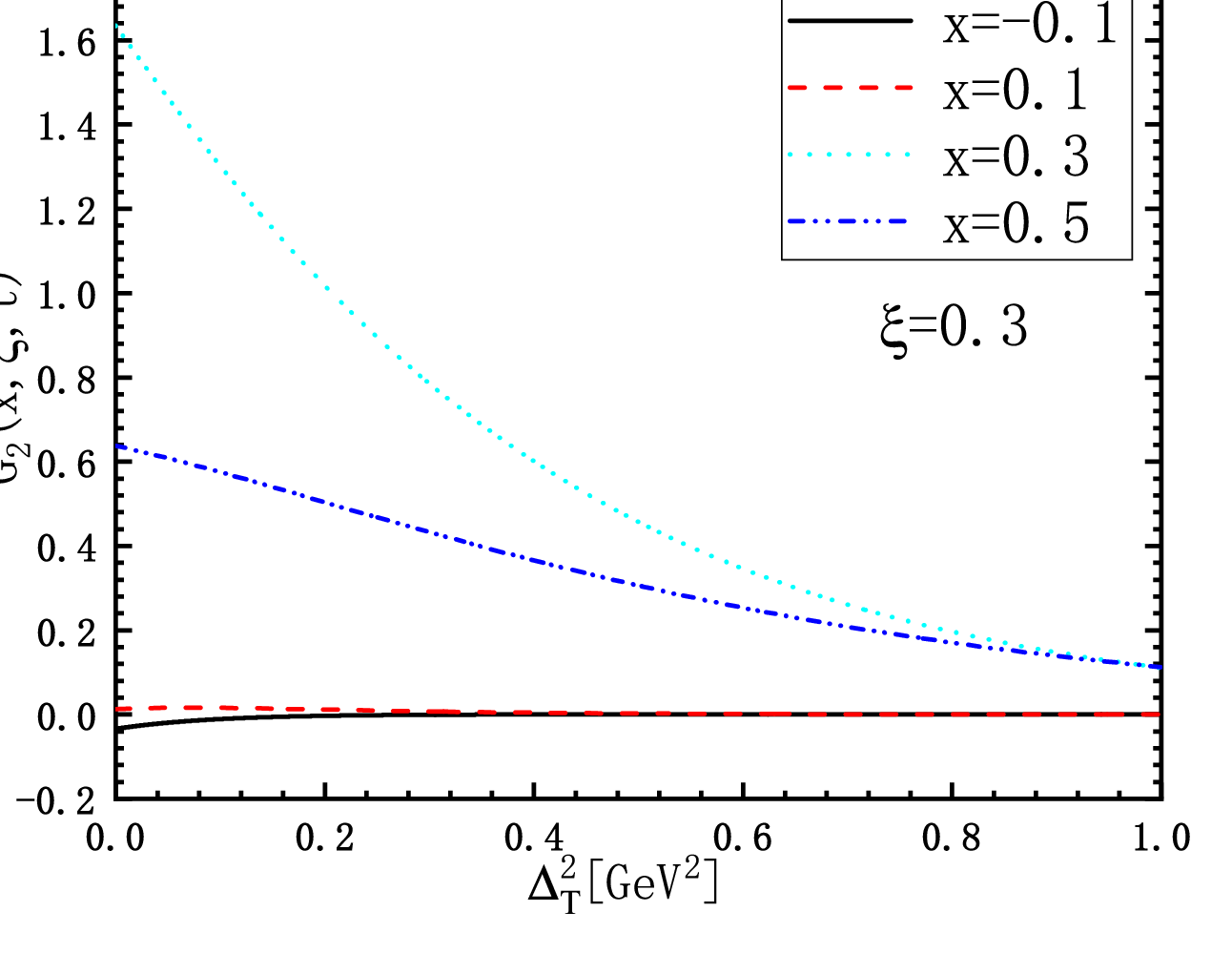}\\
	\includegraphics[width=0.40\columnwidth]{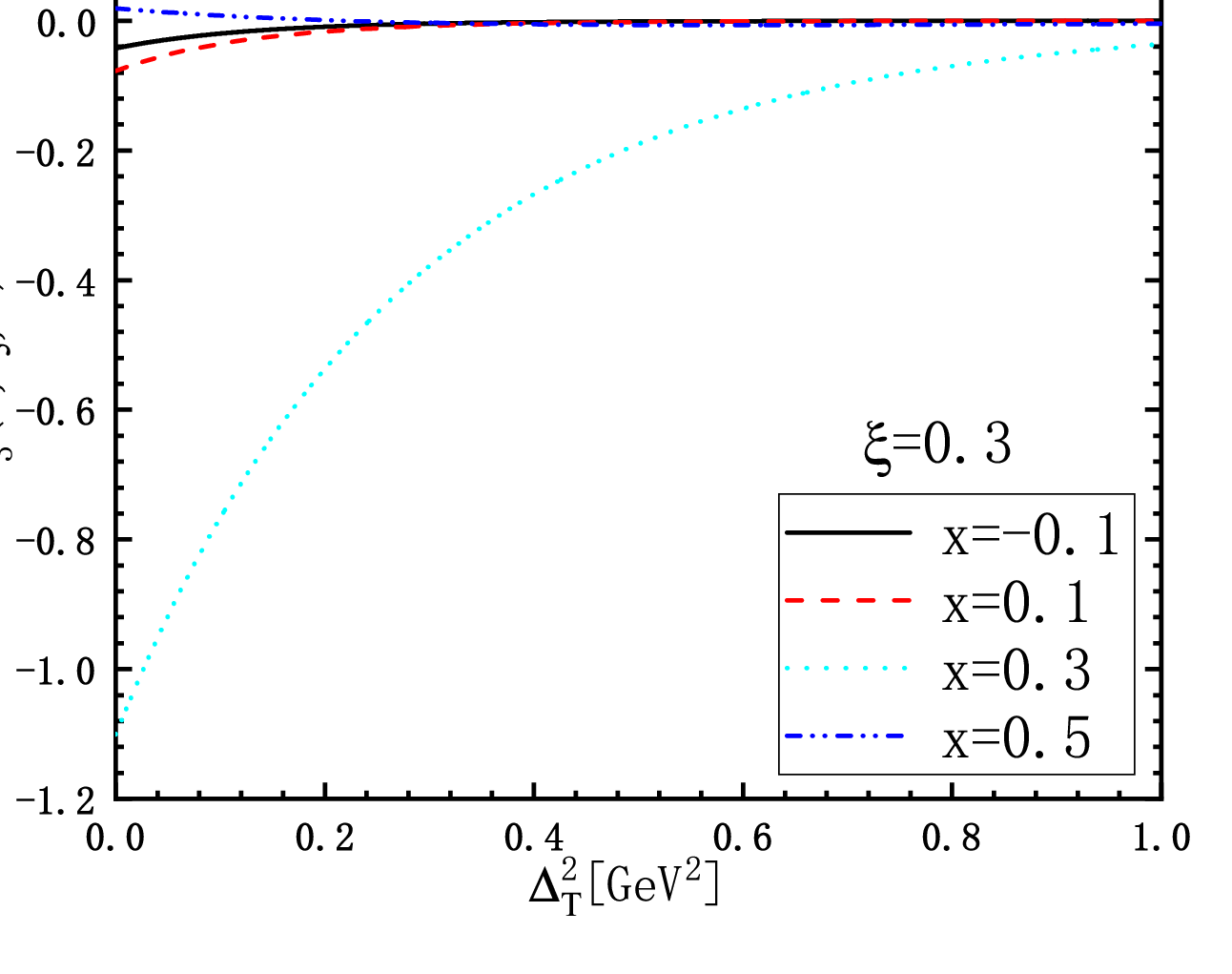}
	\includegraphics[width=0.40\columnwidth]{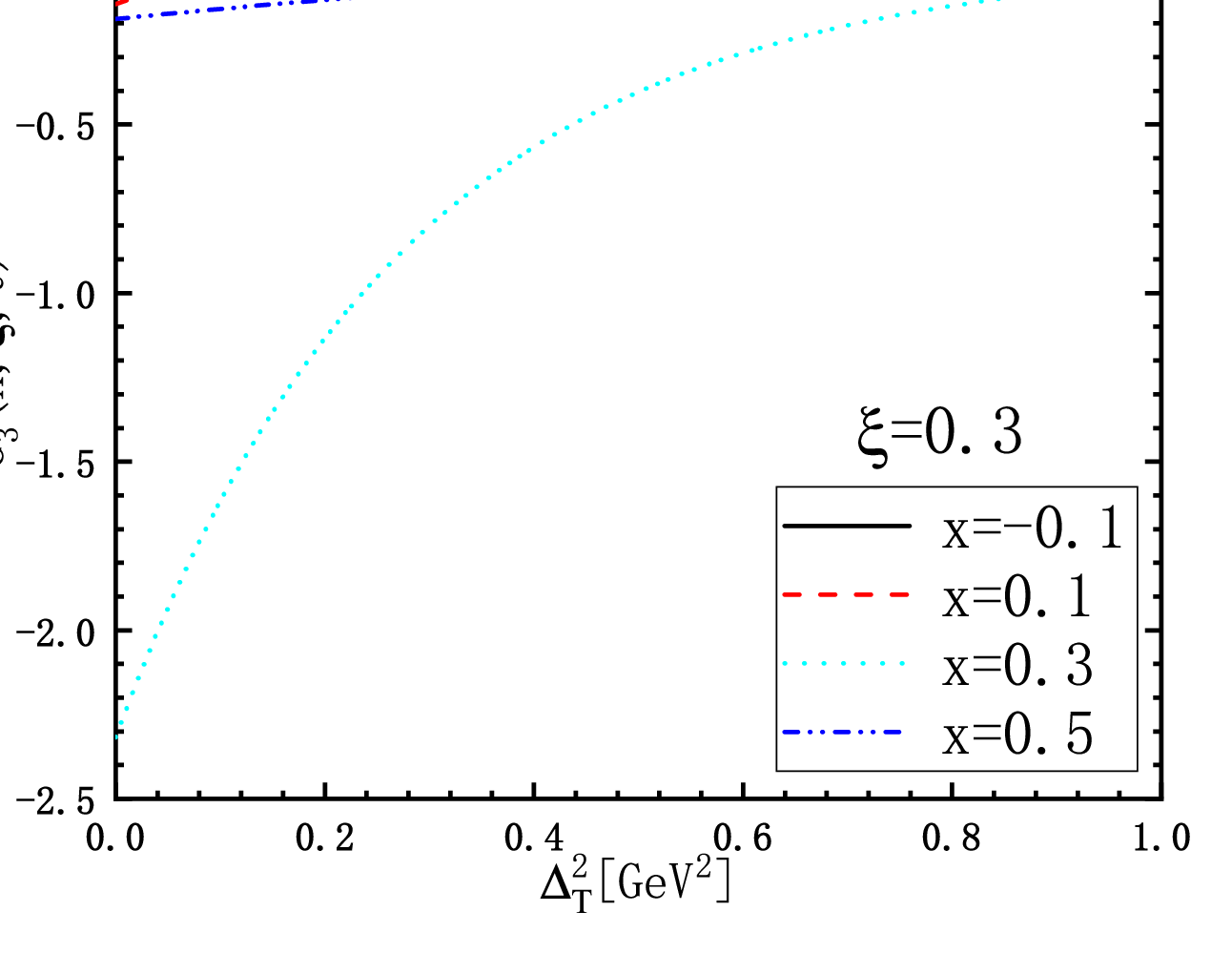}\\
	\includegraphics[width=0.40\columnwidth]{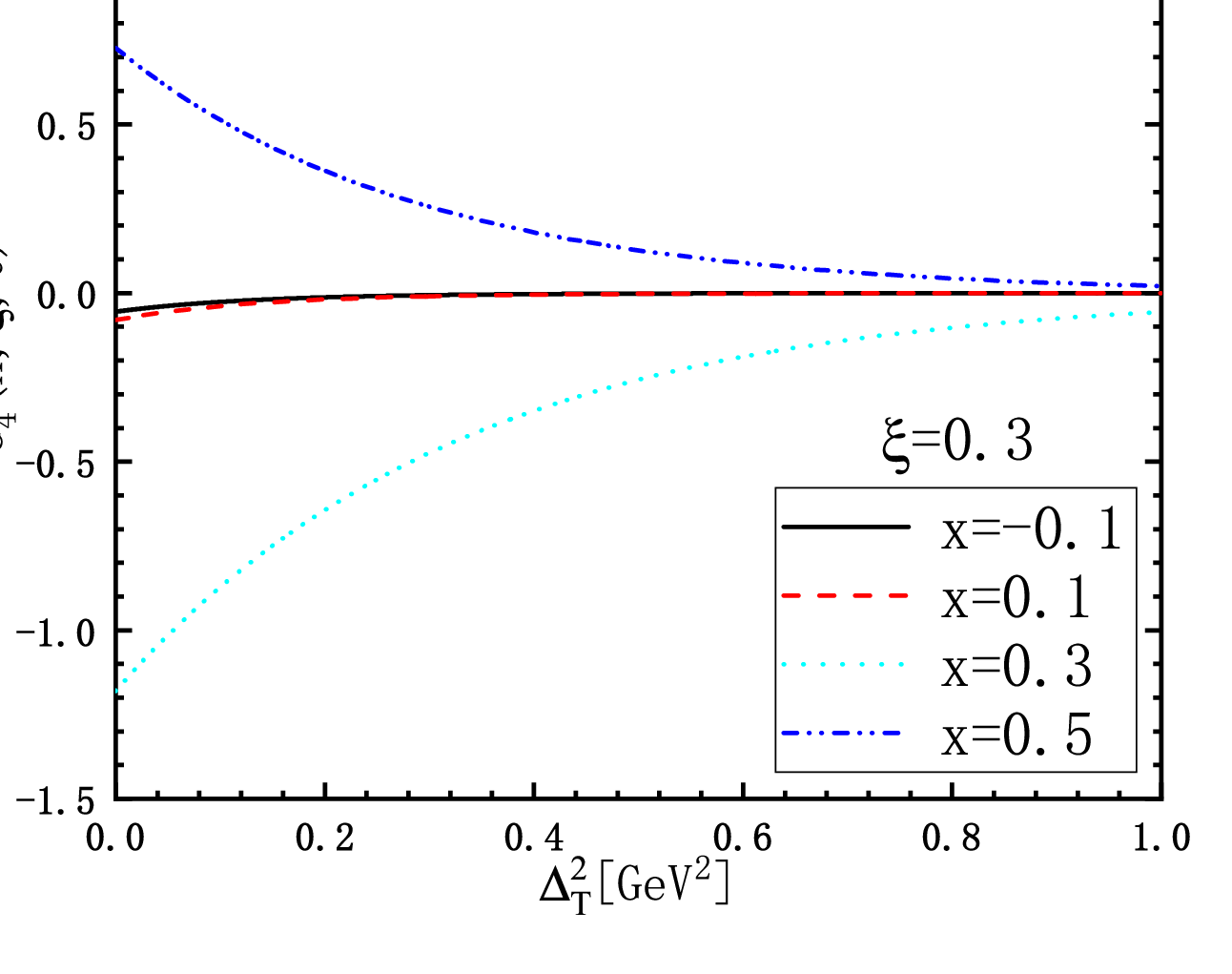}
	\includegraphics[width=0.40\columnwidth]{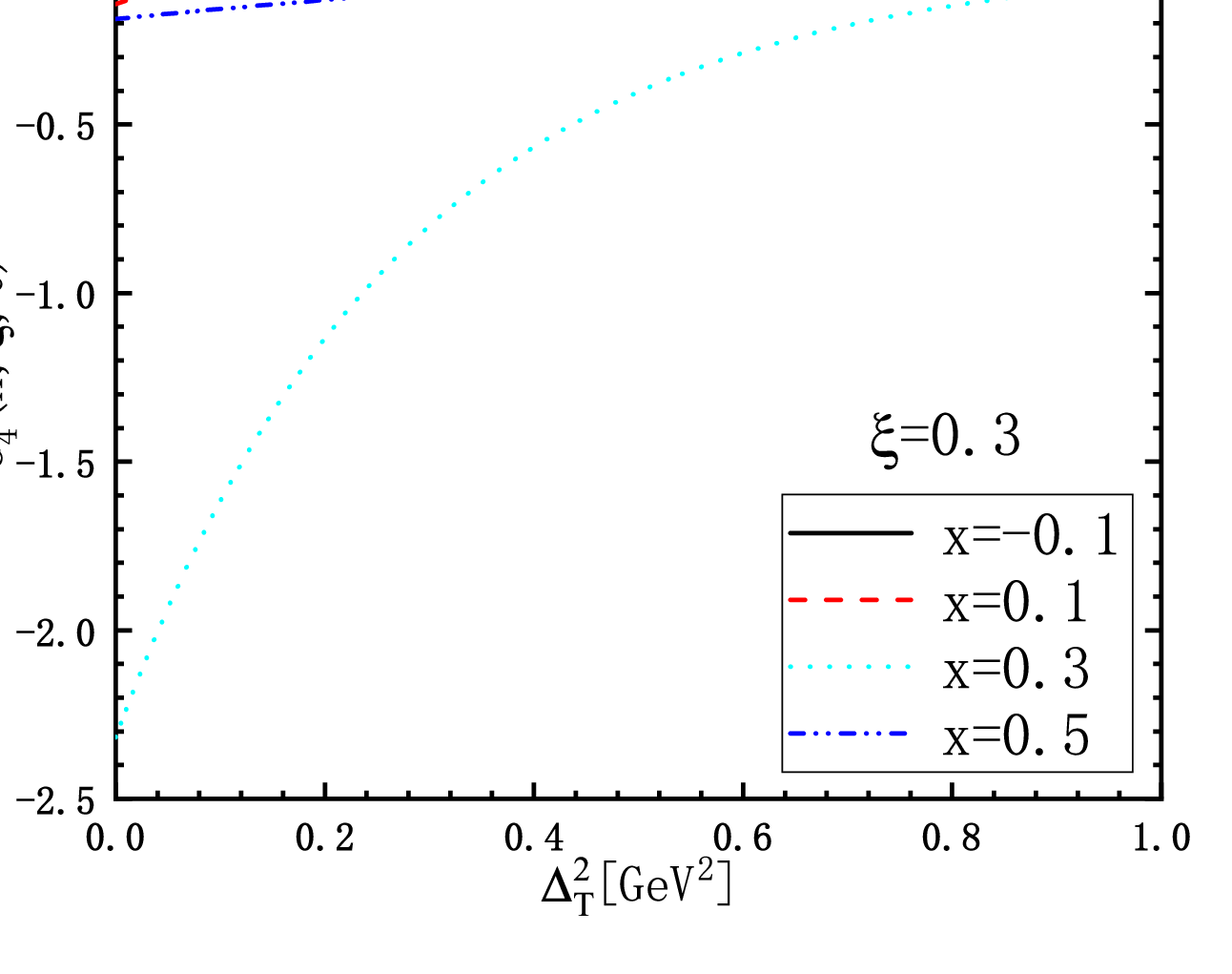}\\
	\caption{The axial-vector twist-3 GPDs $\tilde{G}_1(x,\xi,t)$, $\tilde{G}_2(x,\xi,t)$, $\tilde{G}_3(x,\xi,t)$ and $\tilde{G}_4(x,\xi,t)$ of the $u$ (left panel) and $d$ (right panel) quarks as functions of $\bm{\Delta}_T^2$ at fixed $\xi=0.3$ for $x=-0.1$, 0.1, 0.3, 0.5, respectively.}
	\label{fig:axialvectorGPDdelta}
\end{figure}

In Fig.~(\ref{fig:vectorGPDx}), we plot the vector twist-3 GPDs of $u$ and $d$ quarks vs $x$ at different $\xi$, with $|\bm{\Delta}_T|$ fixed as 0.5 GeV. All the GPDs except $G_1$ exhibit significant discontinuities at $x=\pm\xi$. 
Particularly, the curves for $G_2$ show that the discontinuities reverses the sigh of the GPDs from positive values to negative values at $x=\xi>0.1$.
On the contrary, the signs of $G_3$ and $G_4$ are negative for almost entire x region.
The shape of each vector GPD of the $d$ quark is similar to that of the $u$ quark, with a slightly larger size.  
Due to the presence of the factor $\frac{-1}{|x|(1-|x|)}$ in the exponent of the form factor, all distributions converge to 0 as $|x| \to 0$ and $|x|\to 1$. This factor also results in a tendency that the differences between the discontinuous points in the ERBL and DGLAP regions of $G_2$, $G_3$ and $G_4$ approach to 0 as $\xi\rightarrow 0$ and 1. 
For $G_1$, the positions of the peaks are always around $x=0.3$ as $\xi$ increase, with the maximum value occuring at $\xi=0.3$. 
Although $G_1$ remains continuous across the entire $x$ region, its derivative is discontinuous at $x=\pm\xi$.

In Fig.~(\ref{fig:axialvectorGPDx}), we depict the axial-vector twist-3 GPDs of the $u$ and $d$ quarks vs $x$ at different $\xi$, with $|\bm{\Delta}_T|$  fixed as 0.5 GeV. 
We observe that the qualitative behaviors of $\tilde{G}_1$ are similar to those of the $G_1$ except an opposite sign. 
All the GPDs except $\tilde{G}_1$ are discontinuous at $x=\pm\xi$. 
Except for $\tilde{G}_4$, the discontinuity normally does not alter the sign of the GPD.
Again, the shape of each axial-vector GPD of the $d$ quark is similar to that of the $u$ quark, with a slightly larger size. 

In Figs.~(\ref{fig:vectorGPDdelta}) and (\ref{fig:axialvectorGPDdelta}), we present the dependence of the vector and axial-vector twist-3 GPDs on $\bm{\Delta}_T^2$ at $x=-0.1,\ 0.1, \ 0.3, \ 0.5$, with $\xi$ is fixed as 0.3, respectively. 
Given $\xi=0.3$, nonzero results of GPDs should appear in the region of $-0.3\leq x\leq 1$. 
All GPDs display a smooth trend to approach 0 as $\bm{\Delta}_T^2$ increases. 
At $x=0.3$, the GPDs have the largest size.

\subsection{Forward limit}
\begin{figure}
	\centering
	\includegraphics[width=0.40\columnwidth]{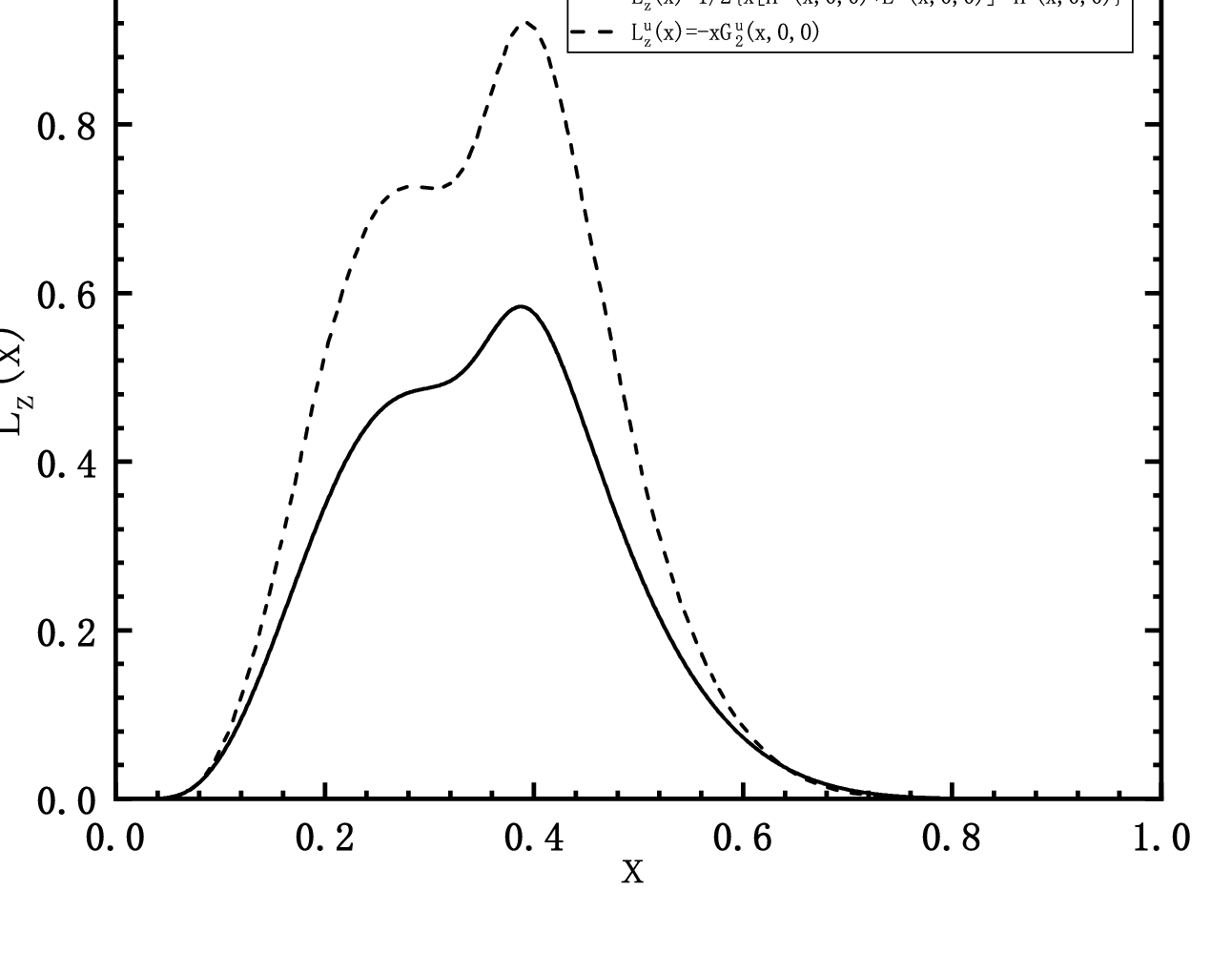}
	\includegraphics[width=0.40\columnwidth]{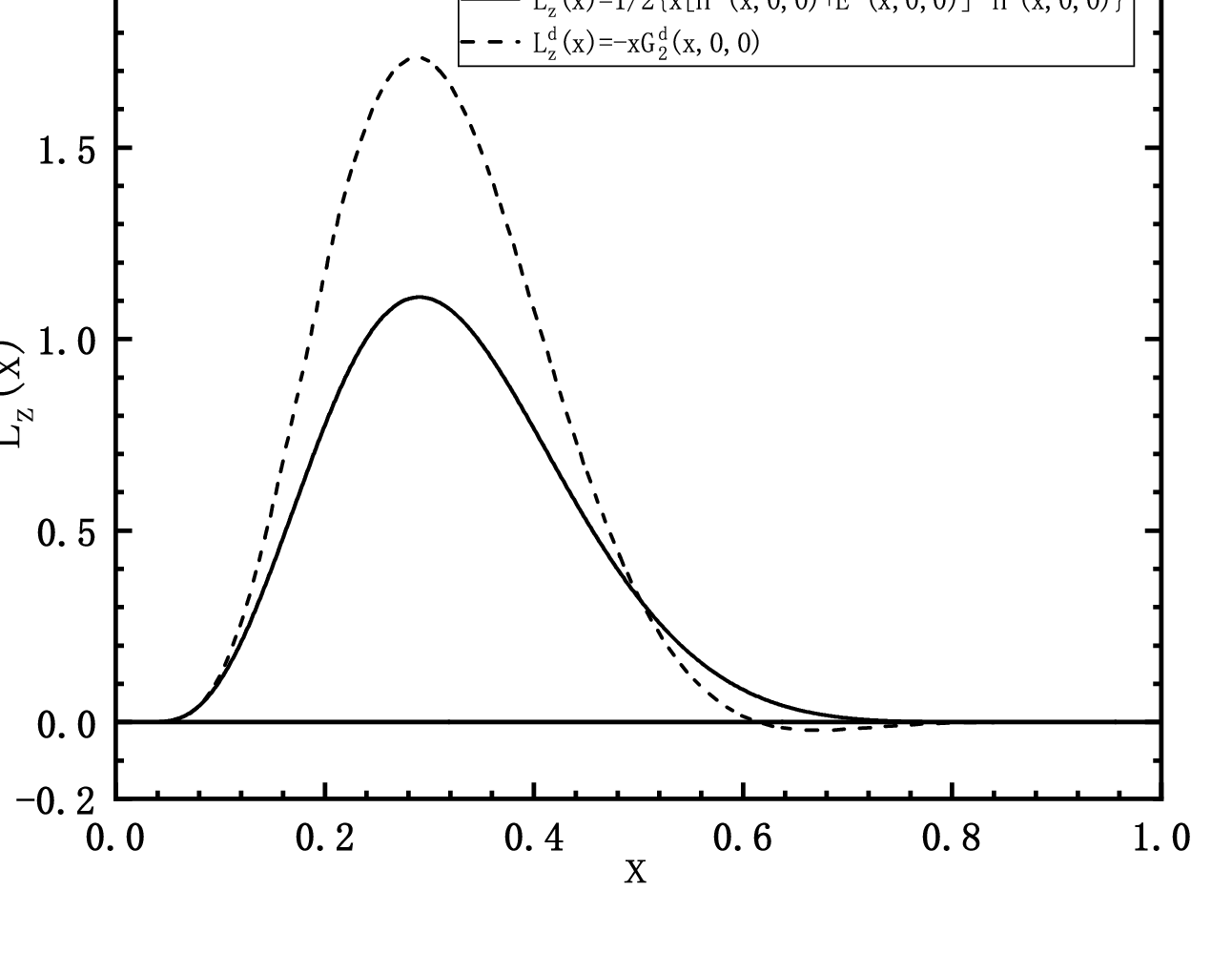}
	\caption{The kinetic OAM of $u$ (left panel) and $d$ (right panel) quarks defined by the twist-2 and twist-3 GPDs as functions of $x$.}
	\label{fig:Lzq}
\end{figure}

Sum rules for the vector and axial-vector twist-3 GPDs $G_i$ and $\tilde{G}_i$ have been listed in Ref.~\cite{Kiptily:2002nx}. For the $x$-moment, there exists a well known sum rule for $G_2$:
\begin{align}
   \int^1_{-1}dxxG_2(x,\xi,t)=\frac{1}{2}\int^1_{-1}dx\{\tilde{H}(x,\xi,t)-x[H(x,\xi,t)+E(x,\xi,t)]\}
   \label{eq:sumrule},
\end{align}
where the right-hand side represents the negative kinetic (or Ji's) OAM of quarks in the proton in the forward limit $\Delta \to 0$~\cite{Ji:1996ek}. 
Alternatively, the kinetic OAM can also be expressed as~\cite{Lorce:2016nxs}
\begin{align}
	L_z^q=-\int^1_{-1}dxxG^q_2(x,0,0)
	\label{eq:Lzq},
\end{align}
which is the chiral-odd analogue of the Penttinen-Polyakov-Shuvaev-Strikman relation~\cite{Kiptily:2002nx,Penttinen:2000dg}. This underscores that the relevance between twist-3 GPDs and physical observables cannot be neglected. 
In Fig.~(\ref{fig:Lzq}), we compare the $x$-dependence of the OAMs of $u$ and $d$ quarks defined by the twist-2 and twist-3 GPDs, respectively. 
It is shown that they have similar shapes, highlighting the similarities of the two approaches.
However, their sizes are different. 
The total OAMs of valence quarks can been obtained by integrating $x$: $L_z^u \approx 0.181$, $L_z^d \approx 0.318$ from twist-2 GPDs, and $L_z^u \approx 0.270$, $L_z^d \approx 0.442$ from twist-3 GPDs.
This indicate that the sum rule is only approximately fulfilled in our model with en error about $30\%$. 
Future refinements, particularly involving higher-order gluonic contributions, are anticipated to improve accuracy.
 
\begin{figure}
	\centering
	\includegraphics[width=0.40\columnwidth]{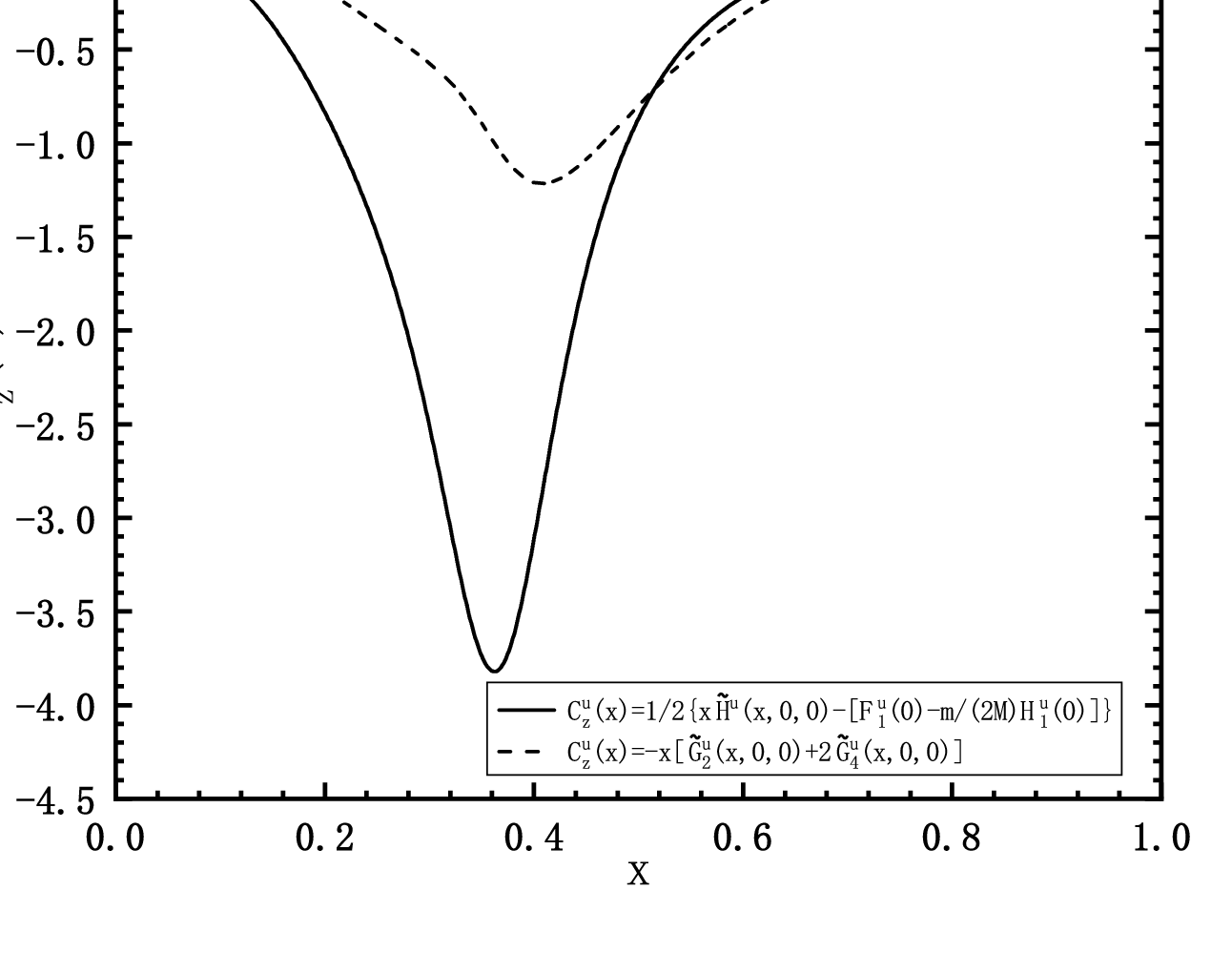}
	\includegraphics[width=0.40\columnwidth]{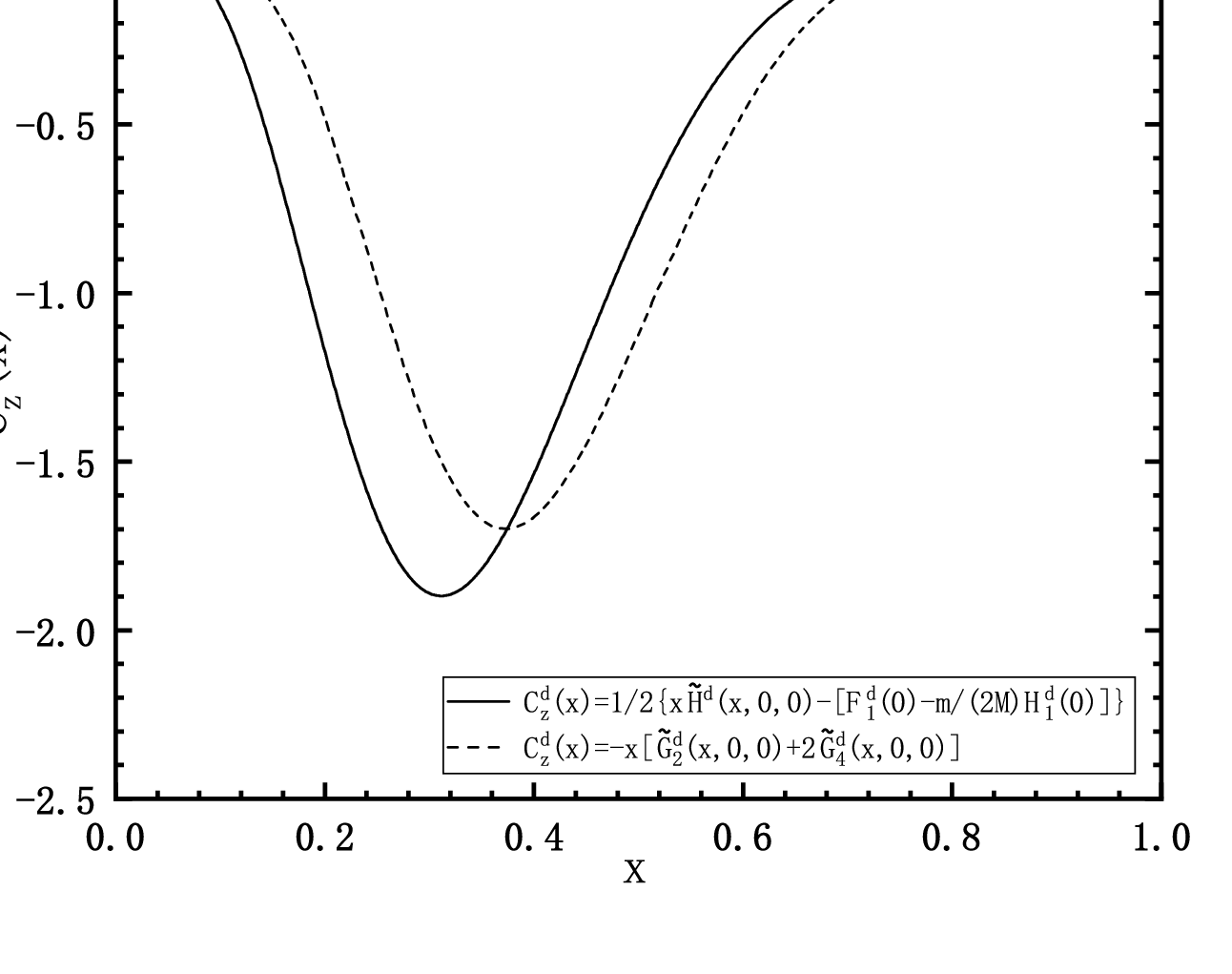}
	\caption{The longitudinal spin-orbit correlations of $u$ (left panel) and $d$ (right panel) quarks defined by the twist-2 and twist-3 GPDs as functions of $x$.}
	\label{fig:Czq}
\end{figure}

Based on the forward limits of twist-3 GPDs $\tilde{G}_2$ and $\tilde{G}_4$, one can provide the chiral-odd analogue of the relation derived in Ref.~\cite{Lorce:2014mxa} for the quark longitudinal spin-orbit correlations~\cite{Bhoonah:2017olu}:
\begin{align}
	C_z^q=-\int^1_{-1}dxx[\tilde{G}^q_2(x,0,0)+2\tilde{G}_4^q(x,0,0)],
\end{align}
which can also be regarded as the analogue of Eq.~(\ref{eq:Lzq}) in the parity-odd sector. In Fig.~(\ref{fig:Czq}), we once again compare the $x$-dependence of the longitudinal spin-orbit correlations of $u$ and $d$ quarks defined by the twist-2 and twist-3 GPDs, respectively. 
Both definitions lead to fairly similar results, but the corresponding sum rule has been broken. 
We observe that the correlations of valence quarks are negative throughout the entire $x$ region, with the main contributions concentrated in $x < 0.5$. 
Compared with other model calculations in Ref.~\cite{Lorce:2014mxa}, we find that more reliable total correlations can be obtained using twist-2 GPDs: $C_z^u = -0.775$, $C_z^d = -0.586$, 
which indicates that the quark spin and OAM in the proton tend to be antialigned, with the correlation of $u$ quarks stronger than that of $d$ quarks. 
These results also imply that higher-order corrections should be included in calculating the higher-twist GPDs in order to provide more accurate results, especially since they are usually measurable at relatively low $Q^2$.

\begin{figure}
	\centering
	\includegraphics[width=0.40\columnwidth]{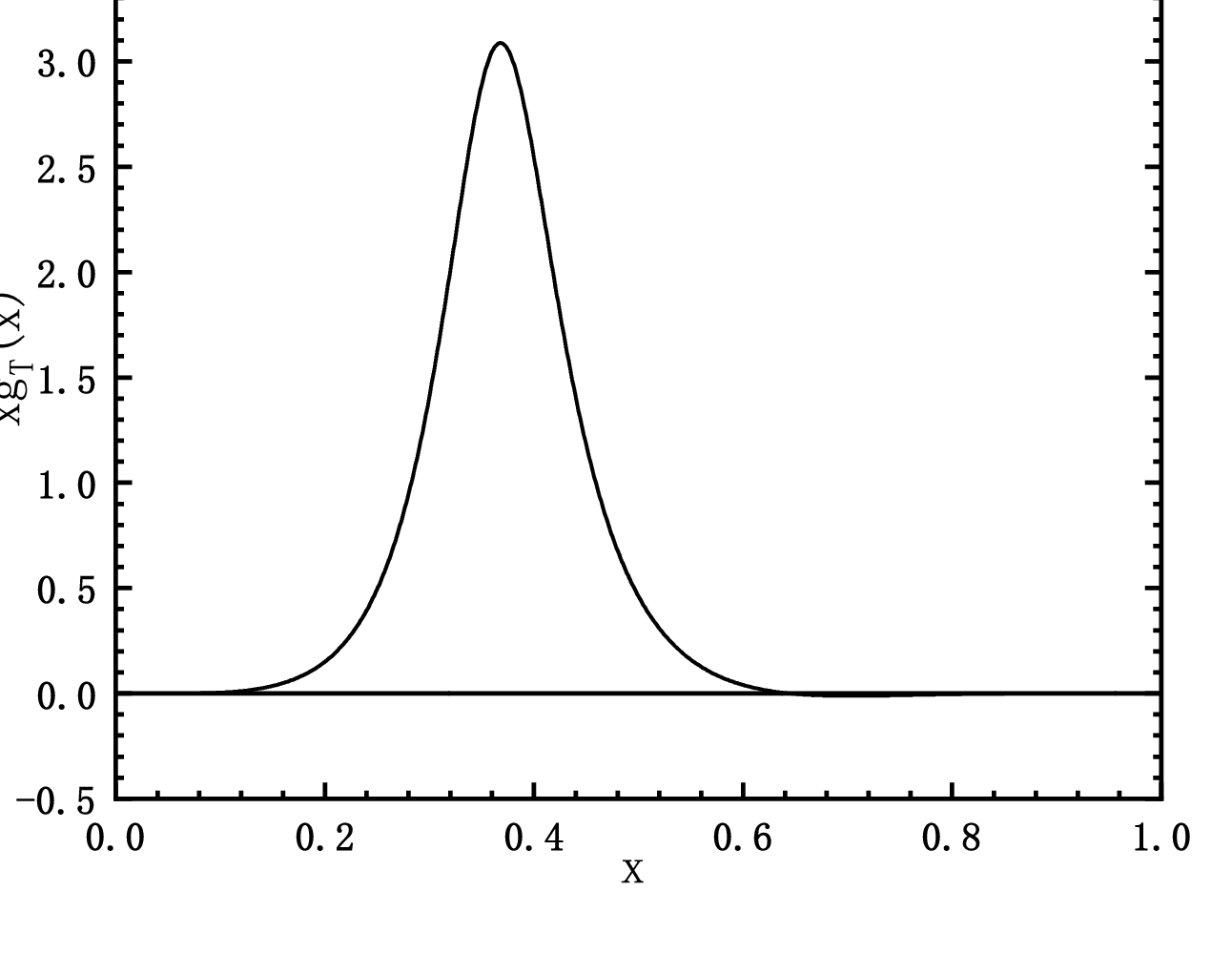}
	\includegraphics[width=0.40\columnwidth]{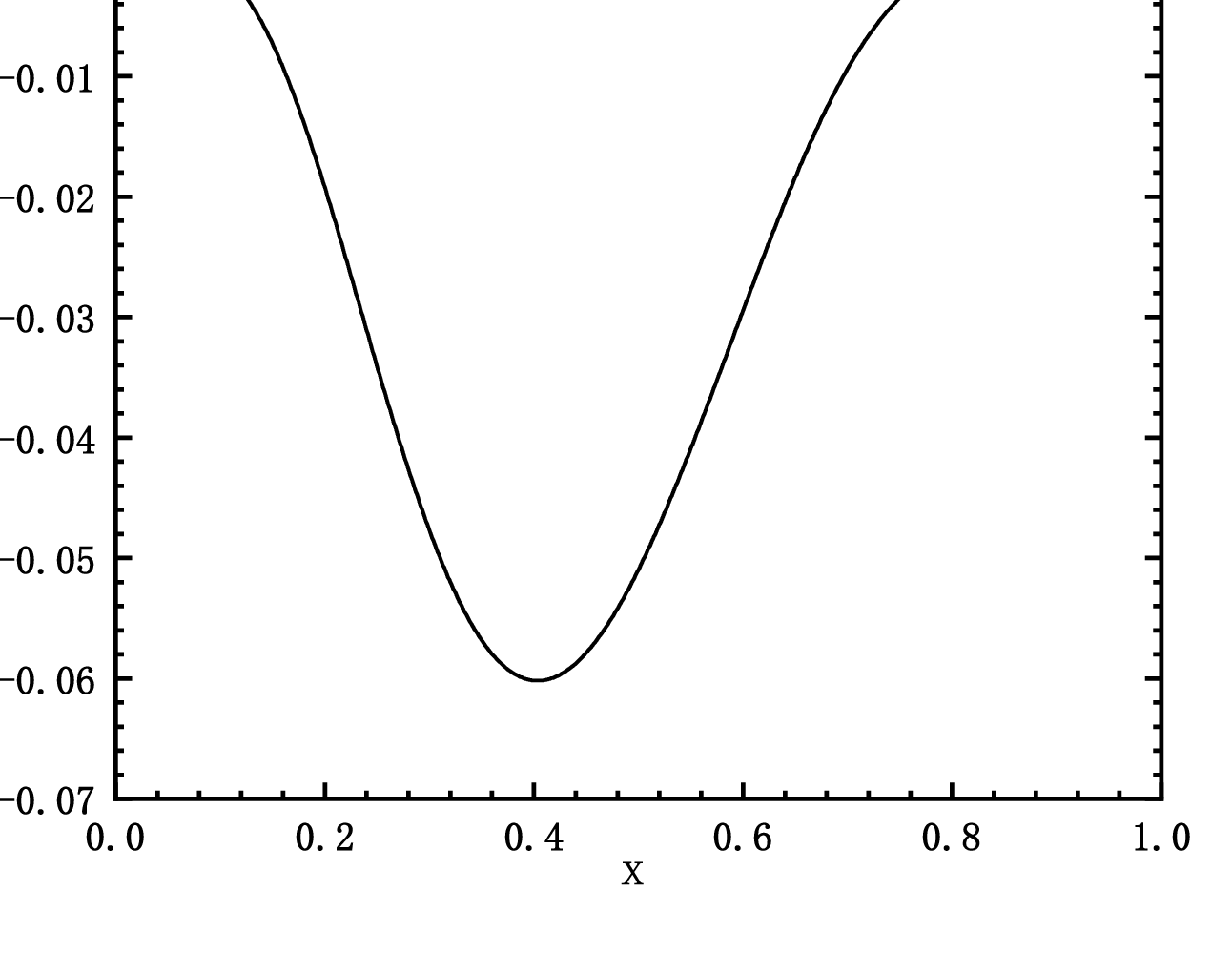}
	\caption{The transverse spin distributions $g_T(x)$ (timed with $x$) of $u$ (left panel) and $d$ (right panel) quarks as functions of $x$.}
	\label{fig:gT}
\end{figure}

Among the chiral-even twist-3 GPDs, only $H_{2T}^\prime(x,\xi,t)$ has a corresponding PDF limit,
\begin{align}
	g_T(x)=\lim\limits_{\Delta \to 0} H_{2T}^\prime(x,\xi,t).
\end{align}
This twist-3 PDF $g_T(x)$, although having no partonic probability interpretation, is sensitive to the quark-gluon correlation~\cite{Jaffe:1991ra}.
The twist-2 PDF $g_1(x)$, parameterized by a similar $\Gamma$ structure, describes the quark helicity distribution in a longitudinally polarized nucleon. 
As shown in Fig.~(\ref{fig:gT}), the magnitude of $g_T^u(x)$ is much larger than that of $g_T^d(x)$, indicating that the contribution from the scalar isoscalar (quark $u$ with diquark $s$) configuration dominates in the proton. 
In addition, the Lorentz invariance of twist-3 PDFs induces the so-called Burkhardt-Cottingham sum rule~\cite{Burkhardt:1970ti}
\begin{align} \int^1_{-1}dxg_T(x)=\int^1_{-1}dxg_1(x),~~~~~~~~\text{or}~~~~~~~~~~\int^1_{-1}dxg_2(x)=0,
	\label{eq:BCsumrule}
\end{align}
where $g_2(x)=g_T(x)-g_1(x)$. This sum rule indicates that the quark spin makes the same contribution to the nucleon spin in different polarizations. 
We have numerically checked this sum rule with our model. 
We find that  our model result for $d$ quark approximately satisfies the sum rule,  however, that for $u$ quark breaks this sum rule.

\section{Conclusion}\label{Sec:5}

In this paper, we applied the spectator diquark model to study the chiral-even twist-3 GPDs for the $u$ and $d$ quarks in the proton with $\xi \neq 0$.
We considered two types of parameterizations for the GPDs and presented the numerical results of the vector and axial-vector twist-3 GPDs. 
The form factor of the nucleon-quark-diquark vertex was chosen as exponential to eliminate the divergence caused by integrating $\bm{k}_T$ in the correlators. For the axial-vector diquark, we adopted the polarization sum which only contains the light-cone transverse polarization states of the axial-vector diquark. 
We found that the GPDs $G_i$ and $\tilde{G}_i$ calculated with the scalar diquarks share similar qualitative features with those calculated with the axial-vector diquarks.
The twist-3 GPDs exhibit discontinuities at $x=\pm\xi$, except for $G_1$ and $\tilde{G}_1$ (or $\tilde{H}_{2T}$ and $\tilde{H}_{2T}^\prime$). 
These discontinuities disappear as $\xi \rightarrow 0$. 
In addition, all GPDs approach 0 as $\bm{\Delta}_T^2$ increases. 

The property of the forward limits of several twist-3 GPDs were explored and several features were found.  
Firstly, due to considering only the lowest order of the process, all sum rules for twist-3 GPDs have been broken. We hope that these can be improved after higher-order contributions are involved. 
Secondly, the negative spin-orbit correlations indicate that the spin and OAM of valence quarks in the proton tend to be antialigned. 
Thirdly, the main contribution to the twist-3 distribution $g_T$ for quarks in the proton comes from the scalar isoscalar (quark $u$ with diquark $s$) configuration. 

The advantage of studying GPDs at $\xi \neq 0$ is that they can be directly connected to the DVCS cross section, and this hard exclusive process is expected to be measured at the EIC and EicC. Further investigate on the higher-order contributions to twist-3 GPDs may  help to reevaluate the related sum rules and deepen our understanding of the proton structure.

\section*{Acknowledgements}
This work is partially supported by the National Natural Science Foundation of China under grant number 12150013.

\end{document}